\documentclass[conference]{IEEEtran}
\IEEEoverridecommandlockouts

\usepackage{multirow}
\usepackage{subcaption}
\usepackage{xcolor}
\usepackage{graphicx}
\usepackage{amsmath}

\def\BibTeX{{\rm B\kern-.05em{\sc i\kern-.025em b}\kern-.08em
    T\kern-.1667em\lower.7ex\hbox{E}\kern-.125emX}}
\begin{document}

\title{TransECG: Leveraging Transformers for Explainable ECG Re-identification Risk Analysis}

\author{
\IEEEauthorblockN{
Ziyu Wang\textsuperscript{*1}, Elahe Khatibi\textsuperscript{*1}, Kianoosh Kazemi\textsuperscript{2}, \\ 
Iman Azimi\textsuperscript{1}, Sanaz Mousavi\textsuperscript{3}, Shaista Malik\textsuperscript{1}, Amir M. Rahmani\textsuperscript{1}
}
\IEEEauthorblockA{
\textsuperscript{1}University of California, Irvine, Irvine, CA, USA \\
\textsuperscript{2}University of Turku, Turku, Finland \\
\textsuperscript{3}California State University, Dominguez Hills, Carson, CA, USA
}
\IEEEauthorblockA{
\textsuperscript{*}Equal contribution. Co-first authors. \\
\{ziyuw31, ekhatibi, azimii, smalik, amirr1\}@uci.edu, \\ 
kianoosh.k.kazemi@utu.fi, srahimimoosavi@csudh.edu
}
}

\maketitle


\definecolor{todo}{RGB}{255,0,0}
\definecolor{elahe}{RGB}{128,128,128}
\definecolor{ziyu}{RGB}{139,0,0}

\newcommand{\todo}[1]{\textcolor{todo}{\hl{#1}}}
\newcommand{\elahe}[1]{\textcolor{elahe}{\hl{#1}}}
\newcommand{\ziyu}[1]{\textcolor{ziyu}{\hl{#1}}}
\newcommand{\thickhline}{\noalign{\hrule height 1pt}}

\begin{abstract}
Electrocardiogram (ECG) signals are widely shared across multiple clinical applications for diagnosis, health monitoring, and biometric authentication. While valuable for healthcare, they also carry unique biometric identifiers that pose privacy risks, especially when ECG data shared across multiple entities. These risks are amplified in shared environments, where re-identification threats can compromise patient privacy. Existing deep learning re-identification models prioritize accuracy but lack explainability, making it challenging to understand how the unique biometric characteristics encoded within ECG signals are recognized and utilized for identification. Without these insights, despite high accuracy, developing secure and trustable ECG data-sharing frameworks remains difficult, especially in diverse, multi-source environments. In this work, we introduce TransECG, a Vision Transformer (ViT)-based method that uses attention mechanisms to pinpoint critical ECG segments associated with re-identification tasks like gender, age, and participant ID. Our approach demonstrates high accuracy (89.9\% for gender, 89.9\% for age, and 88.6\% for ID re-identification) across four real-world datasets with 87 participants. Importantly, we provide key insights into ECG components such as the R-wave, QRS complex, and P-Q interval in re-identification. For example, in the gender classification, the R wave contributed 58.29\% to the model's attention, while in the age classification, the P-R interval contributed 46.29\%. By combining high predictive performance with enhanced explainability, TransECG provides a robust solution for privacy-conscious ECG data sharing, supporting the development of secure and trusted healthcare data environment.
\end{abstract}

\section{Introduction}

Integrating biosignal data, particularly Electrocardiogram (ECG) signals, have significantly advanced the diagnosis and treatment of cardiovascular diseases. ECGs evaluate the electrical activity of the heart, enabling clinicians to identify heart failure, arrhythmias, cardiomegaly, and various cardiac conditions~\cite{van2004clinical}. The physiological variations in ECG segments—such as differences in the P-Q interval, QRS complex, and S-T segment—have been well-documented across gender and age groups ~\cite{surawicz2003differences, wei2019physiology}, making these signals valuable not only for clinical diagnosis~\cite{alikhani2024seal, alikhani2024ea, aqajari2024enhancing} but also for biometric identification~\cite{odinaka2012ecg, ashley2004cardiology}. Such features have been effectively leveraged for demographic classification and biometric re-identification ~\cite{aydin2016effects}.

Re-identification methods are crucial for assessing privacy risks in biometric data, including ECG signals, by identifying whether anonymized datasets can reveal individual identities ~\cite{ghazarian2022assessing}. Techniques like linkage and membership inference attacks highlight vulnerabilities in data-sharing frameworks ~\cite{shokri2017membership}. These methods are essential for mitigating privacy risks, particularly in healthcare, where patient confidentiality is critical. By evaluating such risks, re-identification methods guide the development of privacy-preserving algorithms~\cite{yao2020privacy, yang2022zebra}, such as selective noise addition~\cite{wang2024differential}, ensuring secure and ethical use of ECG data in applications like demographic classification, authentication, and clinical diagnosis.

The widespread adoption of ECG data in healthcare systems, coupled with rapid advances in AI-driven analysis~\cite{siontis2021artificial}, has created a critical tension between data utility and privacy protection~\cite{wang2020guardhealth}. Unique biometric signatures within ECG signals make them vulnerable to re-identification attacks~\cite{ghazarian2022assessing}, including linkage attacks that cross-reference multiple datasets~\cite{kho2015design} and membership inference attacks that expose dataset participation~\cite{shokri2017membership}. These privacy vulnerabilities pose significant challenges for openly sharing ECG data for public research purposes, thereby limiting data availability and hindering progress in AI-driven healthcare research~\cite{tertulino2024privacy}. Moreover, re-identification breaches not only compromise patient confidentiality but also create substantial legal and ethical challenges for healthcare organizations.

While deep learning models, such as CNNs~\cite{labati2019deep} and LSTMs~\cite{jyotishi2021ecg}, have achieved high accuracy in ECG-based identification, they operate as black boxes~\cite{bender2023analysis}, offering minimal insight into which ECG features drive their decisions~\cite{ghazarian2022assessing}. This lack of transparency hinders the ability to assess and mitigate privacy risks, making it challenging for stakeholders to trust and implement these systems in shared healthcare environments. Existing explainable approaches like "ECG Unveiled"~\cite{wang2024ecg}, which rely on manually extracted features and basic models, fall short in addressing the complexities of modern deep learning systems. This gap emphasizes the urgent need for explainable deep learning approaches that can both achieve high accuracy and reveal the patterns most influential in the re-identification process, ensuring secure and privacy-conscious data sharing in healthcare applications~\cite{li2025skewed}.

\begin{figure}[h!]
    \centering
    \includegraphics[width=0.75\linewidth]{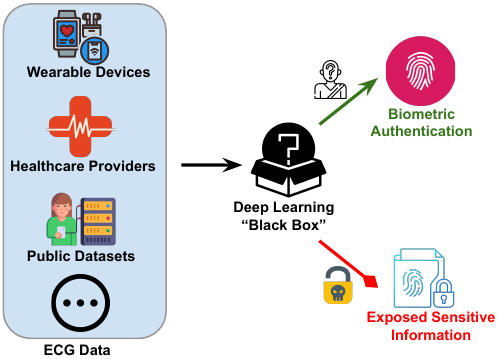}
    \caption{ECG data is widely used in healthcare data spaces for diagnosis, monitoring, and biometric applications but is vulnerable to exploitation. Understanding 'black-box' deep learning models is crucial for secure, privacy-preserving data sharing.}
    \label{fig:motivation}
\end{figure}

In this work, we present TransECG, a Vision Transformer (ViT)-based framework that not only addresses the critical gap between accuracy and explainability in ECG re-identification systems but also advances the development of secure and trusted healthcare data sharing systems.

\textbf{The code for TransECG will be publicly available upon acceptance.}

This work makes the following key contributions:

\begin{itemize}
    \item Developed a ViT-based method for ECG re-identification analysis, leveraging attention mechanisms to identify critical ECG segments that influence identification tasks, enhancing transparency and privacy protection.
    
    \item Utilized attention scores from the ViT model to identify influential ECG features, offering insights into which components drive re-identification without additional explainable AI models.
    
    \item Conducted extensive evaluations across four diverse ECG datasets, validating the method's effectiveness in achieving both high accuracy and interpretability for privacy-sensitive healthcare applications.
\end{itemize}

\section{Background and Related Work}

\textbf{ECG Re-identification and Privacy.} ECG signals have proven to be reliable for biometric identification, as their unique patterns allow for accurate personal identification. Early work using traditional machine learning methods like decision trees and random forests relied on manually extracted features but struggled with scalability and generalization when applied to large, diverse datasets~\cite{camps2006neural}. The rise of deep learning has led to significant advancements in classification accuracy with models like CNNs and LSTMs, which automatically learn features from raw ECG signals~\cite{labati2019deep, jyotishi2021ecg, cheng2024efflex}. However, these models lack transparency, making it difficult to pinpoint the features contributing to re-identification, which complicates the development of effective privacy protection mechanisms. As ECG data is increasingly shared across platforms, there is a growing need to assess the risks of re-identification and to develop techniques that ensure data privacy in biometric applications~\cite{ghazarian2022assessing}. While several studies have highlighted the privacy implications of ECG data in healthcare, effective solutions for identifying and mitigating re-identification risks remain underexplored.

\textbf{Explainable AI (XAI) in Healthcare.} XAI has become an essential tool in healthcare for interpreting the decisions of complex models, particularly in areas such as disease diagnosis and treatment planning~\cite{chaddad2023survey, wang2024healthq}. While methods like saliency maps, attention mechanisms, and SHAP have been widely applied to improve model interpretability in clinical settings, their use for addressing privacy risks in biometric data, particularly ECG signals, remains limited~\cite{adebayo2018sanity, lundberg2017unified}. Few studies have explored how XAI can be used to identify features that lead to re-identification in biometric datasets, such as ECG signals, highlighting a gap in the literature regarding the use of XAI for privacy protection in biosignal data. Notably, ECG Unveiled \cite{wang2024ecg} is one of these studies that attempts to address this gap by relying on manually crafted ECG features, such as P-Q, R-S, and S-T intervals, for re-identification tasks and explainability by using SHAP. While this method enhances interpretability, it remains constrained by its static nature, failing to capture the broader, complex dependencies present in the ECG signal and end up with lower classification accuracy.

\textbf{Transformers in Biosignal Analysis.} Transformer models, powered by self-attention mechanisms, have transformed sequential data processing by capturing long-range dependencies~\cite{chefer2021transformer}. In the context of biosignal analysis, particularly ECG data, transformers are more efficient at modeling temporal relationships compared to traditional recurrent models~\cite{ali2022xai}. However, their application to ECG re-identification risk analysis remains underexplored, especially in terms of interpretability. Although attention mechanisms offer some insight into which parts of the signal are influential, integrating explainability into transformers for privacy-sensitive tasks like ECG re-identification requires further investigation.

\section{Method}

\begin{figure*}[htbp]
    \centering
    \includegraphics[width=0.93\linewidth]{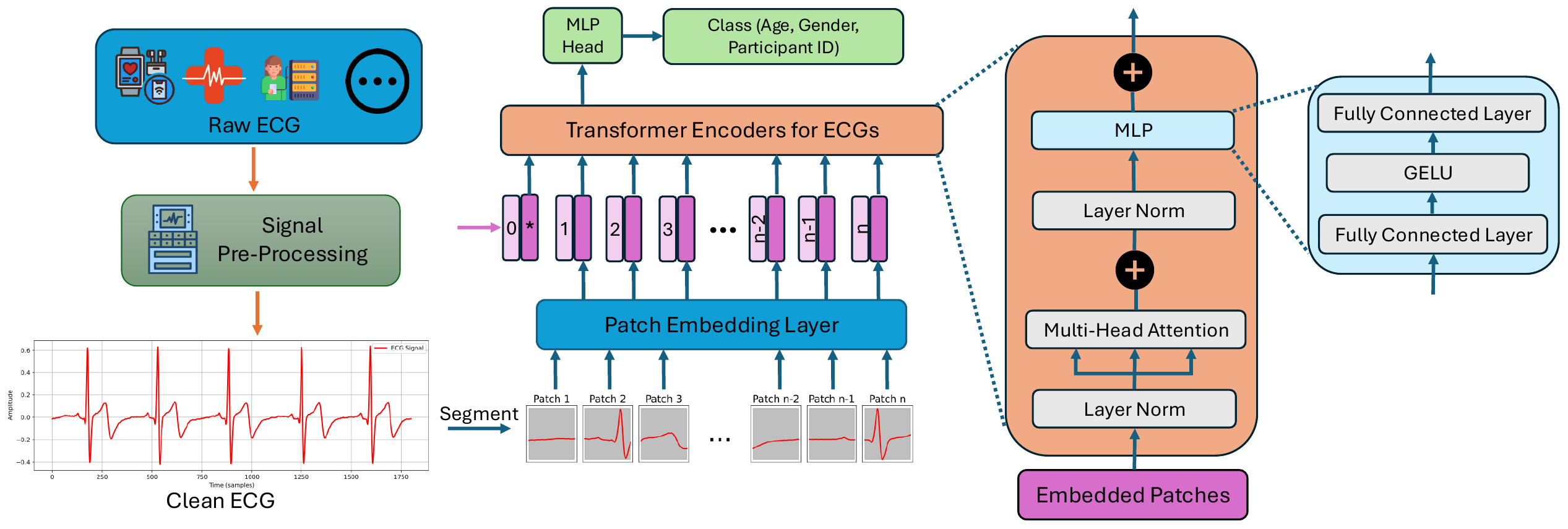}
    \caption{Overview of the proposed TransECG architecture.}
    \label{fig:transECG}
\end{figure*}

We propose a ViT-based method tailored for ECG classification and re-identification risk analysis. The pipeline begins with signal pre-processing to enhance data quality, followed by patch segmentation, embedding, and a transformer-based architecture for feature extraction and classification. 

Fig.~\ref{fig:transECG} illustrates the architecture, which consists of Raw ECG inputs, Signal Pre-Processing, Patch Embedding Layer, Transformer Encoders for ECGs, Multi-Head Attention, and a Multilayer Perceptron (MLP) Head for classification. Each component is designed to handle the specific challenges of ECG analysis, including temporal dependencies, noise, and physiological variability. Integrated explainability techniques provide insights into critical ECG segments for robust decision-making in privacy-sensitive contexts.

\subsection{ViT Architecture for ECG Analysis}
The ViT model~\cite{dosovitskiy2020image} was adapted to process ECG signals, treating them as sequences of patches. This approach captures both local and global dependencies using self-attention mechanisms.

\subsubsection{Signal Pre-Processing}
Signal pre-processing ensures data consistency and compatibility with the model. Raw ECG signals undergo the following steps:
\begin{itemize}
    \item \textbf{Noise Filtering:} A Butterworth bandpass filter removes baseline wander and high-frequency noise, preserving key physiological information such as the P-wave and QRS Complex.
    \item \textbf{Motion Artifact Reduction:} Median filtering addresses motion-induced artifacts caused by physical activity or sensor displacement, ensuring cleaner signals.
    \item \textbf{Normalization:} Signals are resampled to a uniform frequency (e.g., 250 Hz) and normalized to standardize amplitude ranges across recordings, facilitating consistent analysis.
\end{itemize}
This process ensures clean and reliable input for subsequent analysis, improving both data quality and model performance.

\subsubsection{ECG Signal Structure and Patch Embedding}
An ECG signal consists of characteristic segments (P-wave, QRS Complex, T-wave) that represent different phases of the cardiac cycle. These physiological segments vary in length based on cardiac activity and provide essential information about the heart's electrical behavior. To process these signals in a computationally efficient manner, the ECG signal is divided into fixed-length computational units called patches.

These patches differ from clinical ECG segments—while clinical segments correspond to specific cardiac events, the patches used in our model are uniform, non-overlapping divisions of the signal optimized for machine learning tasks. Each patch may contain portions of one or more physiological ECG segments, enabling the model to capture local temporal patterns systematically.

Once divided into patches, each patch is flattened into a vector representation. These flattened patches are then transformed into a $D$-dimensional embedding space using the \textbf{Patch Embedding Layer}, a fully connected layer. The transformation is defined as:

\[
\mathbf{z}_i = \text{Flatten}(\mathbf{x}_i) \cdot \mathbf{E}, \quad \text{for } i = 1, \dots, N,
\]

where $\mathbf{x}_i$ represents the raw data of the $i$-th patch, and $\mathbf{E}$ is the projection matrix. This layer ensures that each patch is mapped into a high-dimensional feature space, enabling the model to process both local and global signal characteristics. Learnable positional embeddings $\mathbf{E}_{\text{pos}}$ are added to the patch embeddings to encode temporal order. A class token $\mathbf{z}_0$ is prepended to the sequence to serve as a global representation for classification. The final input to the transformer encoder is given by:

\[
\mathbf{Z}_0 = [\mathbf{z}_0; \mathbf{z}_1; \mathbf{z}_2; \dots; \mathbf{z}_N] + \mathbf{E}_{\text{pos}}.
\]

\subsubsection{Transformer Encoders for ECGs}
The Transformer Encoders consist of $L$ layers, each comprising Multi-Head Attention (MHSA) and a feedforward neural network (FFN). For layer $\ell$, the computations are:

\begin{align}
\mathbf{Z}'_\ell &= \text{LN} \left( \mathbf{Z}_{\ell - 1} + \text{MHSA}(\mathbf{Z}_{\ell - 1}) \right), \\
\mathbf{Z}_\ell &= \text{LN} \left( \mathbf{Z}'_\ell + \text{FFN}(\mathbf{Z}'_\ell) \right),
\end{align}

where $\text{LN}$ denotes layer normalization. MHSA computes self-attention scores to model relationships between patches. For each attention head $h$:

\[
\mathbf{A}_h = \text{softmax}\left( \frac{\mathbf{Q}_h \mathbf{K}_h^\top} {\sqrt{D_h}} \right) \mathbf{V}_h,
\]

where $\mathbf{Q}_h$, $\mathbf{K}_h$, and $\mathbf{V}_h$ are the query, key, and value matrices, and $D_h$ is the dimensionality of each head. By integrating self-attention across multiple layers, the encoder captures both short-term and long-term dependencies within the signal.

\subsubsection{MLP Head}
The output corresponding to the class token $\mathbf{z}_0$ is passed to the \textbf{MLP Head}, a fully connected layer followed by a softmax activation for classification:

\[
\hat{\mathbf{y}} = \text{softmax}\left( \mathbf{W}_{\text{head}} \cdot \mathbf{z}_0 + \mathbf{b}_{\text{head}} \right),
\]

where $\mathbf{W}_{\text{head}}$ and $\mathbf{b}_{\text{head}}$ are the weights and biases, respectively, and $K$ is the number of classes. This step enables final task-specific predictions, such as gender classification or participant identification.

\subsection{Training and Optimization}
Our framework addresses three ECG classification tasks: gender, age group, and participant ID re-identification. These tasks range from binary (gender) to fine-grained multi-class classification (participant ID). We employ a categorical cross-entropy loss:

\[
\mathcal{L} = - \frac{1}{N} \sum_{i=1}^{N} \sum_{k=1}^{K} y_{i,k} \log \hat{y}_{i,k},
\]

where $y_{i,k}$ and $\hat{y}_{i,k}$ are the ground-truth and predicted probabilities for class $k$. This loss function effectively balances gradients across tasks, enabling simultaneous training. We use Adam optimizer with a learning rate scheduler to adaptively adjust learning rates during training, improving convergence and model performance.

\subsection{Explainability Integration}
The ViT's self-attention mechanism provides inherent explainability by quantifying the importance of ECG patches in classification decisions. Attention scores are extracted from the final transformer encoder block and mapped back to the ECG signal. For attention head $h$, the importance of each patch is computed as:

\[
\mathbf{a}_h = \mathbf{A}_h[0, 1:N],
\]

where $\mathbf{A}_h$ represents the attention scores between the class token and patch embeddings. The overall importance score for patch $i$ is averaged across all heads:

\[
\text{Importance Score}_i = \frac{1}{H} \sum_{h=1}^{H} a_{h,i},
\]

for $i = 1, \dots, N$. These scores are mapped to physiological components (e.g., P-wave, QRS Complex) to reveal which segments most influence predictions. For example, high attention to the R-wave during participant ID re-identification underscores its utility as a biometric marker. This approach transforms the model from a black-box predictor into an interpretable tool, supporting clinical insights and privacy-sensitive applications.

\section{Experimental Results}

In this section, we present the experimental results to validate the performance of TransECG across three classification tasks: gender, age group, and participant ID re-identification. By leveraging attention mechanisms in the ViT architecture, the model achieved high accuracy and provided interpretable insights into the ECG features driving its decisions. These results will be analyzed from two key perspectives: 1) accuracy assessment, highlighting the performance metrics and ROC across tasks, comparison with the baseline, and 2) explainability, emphasizing the role of attention mechanisms in identifying critical ECG segments such as the R-wave and P-R interval for re-identification tasks.

\subsection{Dataset}
We evaluated TransECG on three tasks: gender classification, age group classification, and participant ID re-identification. Our experiments utilized four established ECG databases: the MIT-BIH Arrhythmia Database~\cite{moody2001impact}, MIT-BIH Long-Term ECG Database~\cite{goldberger2000physiobank}, CHF Congestive Heart Failure Database~\cite{baim1986survival}, and the Brno University of Technology ECG Signal Database~\cite{baim1986survival}. These databases were merged to create a comprehensive dataset of 87 participants, with data sorted by participant ID and sample number to maintain sequence integrity.

Our evaluation framework was designed to assess the model's performance across different demographic patterns and individual characteristics. For age classification, we defined five clinically relevant age groups (0-18, 19-35, 36-50, 51-65, and 66+ years), reflecting distinct life stages with documented ECG pattern variations. The gender classification task evaluated binary differences in cardiac signals, while the participant ID re-identification task assessed the model's ability to capture unique individual ECG characteristics. The final dataset includes a diverse range of ages, genders, and cardiac conditions, enabling comprehensive evaluation of the model's generalization capabilities across different population groups.


\subsection{Experiment Setup}

The implementation pipeline comprises three main stages: data preprocessing, model development, and training protocols. For preprocessing, a Butterworth bandpass filter (0.5–40 Hz) was applied to remove noise and baseline drift, followed by the Pan-Tompkins algorithm for R-peak detection. Participants with fewer than 2,000 samples were excluded to ensure consistent sequence lengths. Each ECG signal was resampled to 250 Hz, segmented into sequences of 2,000 time steps, and normalized using min-max scaling at the sequence level.

The TransECG architecture utilizes a ViT with six transformer layers, six attention heads, and a hidden dimension of 256. ECG sequences were divided into patches of size 20, with the model’s MLP having a dimension of 128. Training was performed using the AdamW optimizer with a learning rate of 1×10\^{-4}, employing a stochastic depth survival probability of 0.8. The model was trained for 45 epochs with early stopping based on validation accuracy. The dataset was split into training (70\%), validation (15\%), and test (15\%) sets, ensuring no participant overlap to evaluate generalization effectively.

The framework was implemented using Python libraries, including TensorFlow, TensorFlow Addons, NumPy, pandas, and scikit-learn, for model development and data handling. ECG signals were processed as 1D patches through a convolutional embedding layer and multi-head attention. For reproducibility, datasets must be organized in specified directories, preprocessed using the provided scripts, and trained with consistent parameter settings. The sequential execution of preprocessing, training, and visualization scripts ensures a streamlined and reproducible workflow.

\subsection{Baseline Method}  
ECG Unveiled~\cite{wang2024ecg} acts as a baseline model for comparison purpose, utilizing interpretable machine learning techniques—including logistic regression, decision trees, and random forests—alongside manually engineered ECG features to assess re-identification risks. SHAP analysis is applied to highlight critical features within the PQRST complex that influence classification tasks such as gender, age group, and participant identification.

\subsection{Accuracy Assessment}


Table~\ref{tab:comparison} presents the performance metrics for all classification tasks, including accuracy, precision, recall, and F1-score, which were evaluated on the test set to ensure generalizability to unseen data. The results highlight the robust performance of TransECG across gender, age, and ID re-identification tasks. Notably, for age classification, the model achieved an accuracy of 0.899, significantly surpassing traditional approaches. In ID re-identification, the model reached an accuracy of 0.886, demonstrating its ability to capture individual-specific ECG patterns effectively. Similarly, gender classification achieved an accuracy of 0.899, validating the model’s versatility across diverse tasks.

\subsubsection{Performance Comparison of TransECG and ECG Unveiled Models}

Compared to baseline models such as "ECG Unveiled" \cite{wang2024ecg}, which relies on SHAP-based analysis with manually crafted features, TransECG excels in both predictive performance and explainability. By leveraging ViT's attention mechanisms, TransECG automatically identifies critical ECG components, such as the QRS complex and R-wave, while consistently outperforming ECG Unveiled across all tasks (Table~\ref{tab:comparison}). For gender classification, TransECG achieved a notable improvement in accuracy (0.899 vs. 0.755) and F1-score (0.899 vs. 0.760). Similarly, in age classification, TransECG reached an accuracy of 0.899 and F1-score of 0.899, significantly outperforming ECG Unveiled (accuracy: 0.671, F1-score: 0.633). For ID re-identification, TransECG achieved an accuracy of 0.886 compared to 0.819 by better leveraging attention mechanisms to refine the identification of critical features. These results demonstrate TransECG’s superiority in capturing task-specific ECG variations and enhancing transparency through explainability, offering a superior choice for privacy-sensitive healthcare applications.


\begin{figure}[htbp]
    \centering
    
    \begin{subfigure}[b]{0.43\textwidth}
        \centering
        \includegraphics[width=\linewidth]{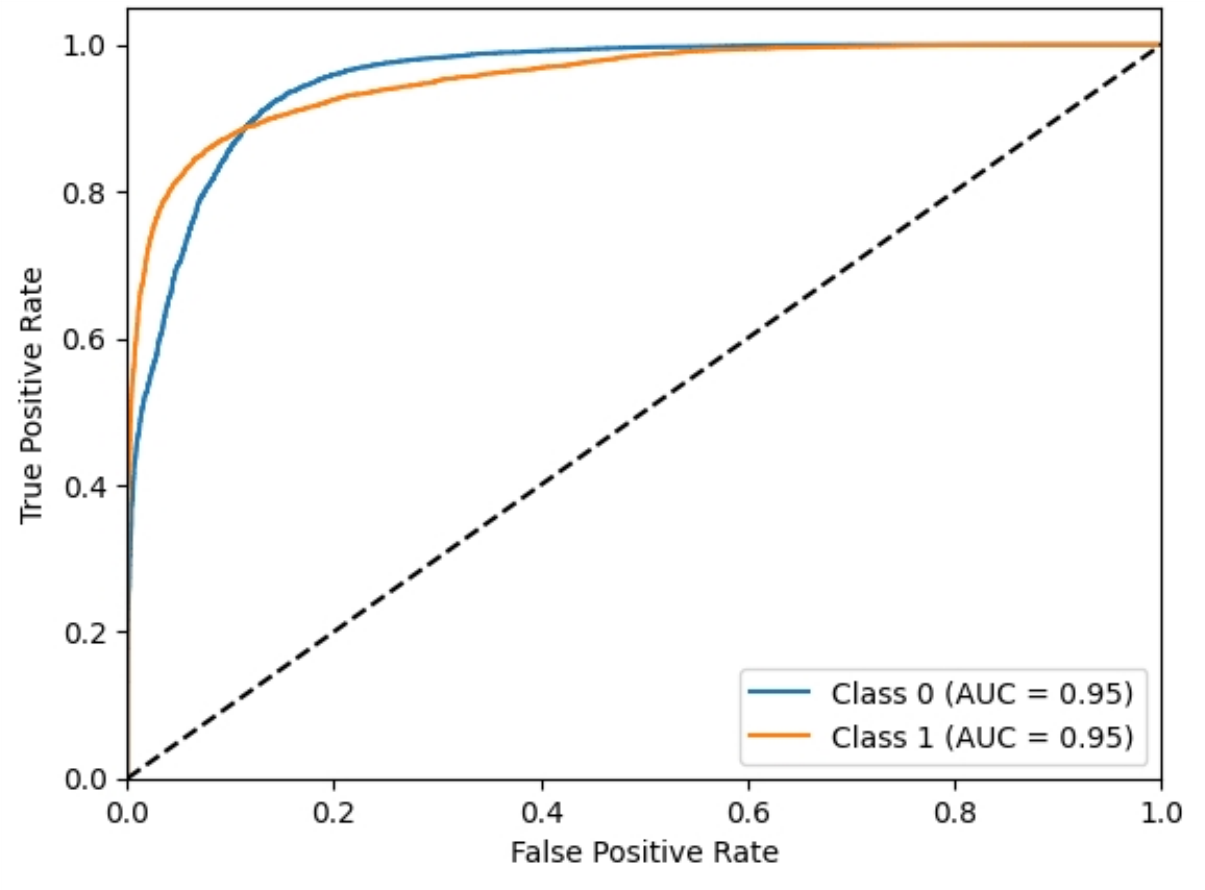} 
        \caption{Multi-Class ROC Curve for Gender Classification}
        \label{fig:roc_gender_classification}
    \end{subfigure}
    \hfill
    \begin{subfigure}[b]{0.43\textwidth}
        \centering
        \includegraphics[width=\linewidth]{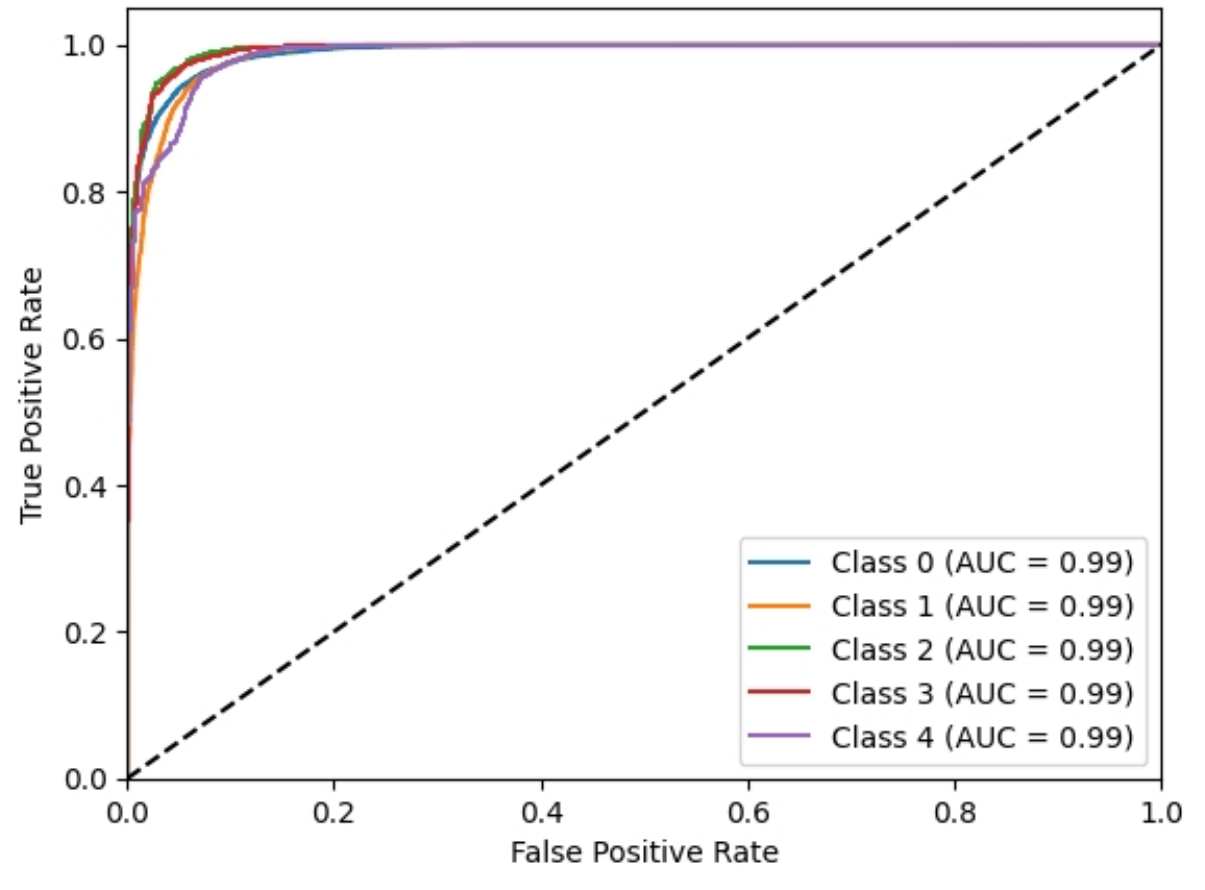} 
        \caption{Multi-Class ROC Curve for Age Classification}
        \label{fig:roc_age_classification}
    \end{subfigure}
    
    \vfill
    \begin{subfigure}[b]{0.43\textwidth}
        \centering
        \includegraphics[width=\linewidth]{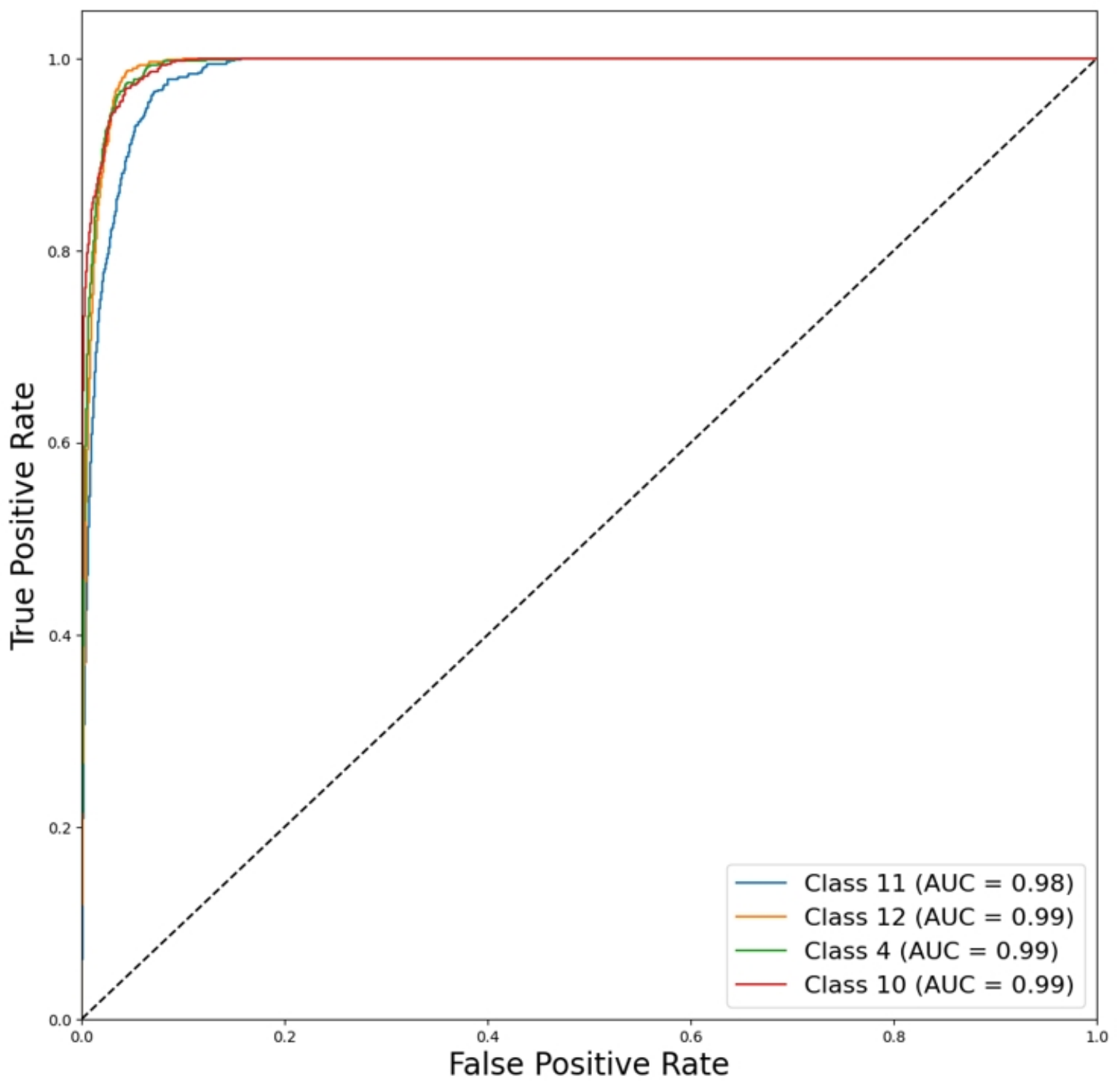} 
        \caption{Multi-Class ROC Curve for ID re-identification}
        \label{fig:roc_id_reidentification}
    \end{subfigure}

    \caption{Comparison of ROC Curves for Three Classification Tasks}
    \label{fig:roc_comparison}
\end{figure}

\subsubsection{ROC Analysis}

To further evaluate the model’s discriminative ability, ROC curves were computed for each task (Fig.~\ref{fig:roc_comparison}). The area under the curve (AUC) values were 0.95 for gender classification, 0.99 for age classification, and 0.99 for ID re-identification (with the top four classes displayed for the latter due to the high number of participants). These high AUC scores highlight the model's effectiveness in distinguishing between classes, particularly in the more challenging tasks of age classification and ID re-identification. These results reinforce the robustness and reliability of TransECG across all tasks.

\begin{table*}[h]
\centering
\caption{Comparison of Performance Metrics between TransECG and Other Baselines}
\footnotesize  
\setlength{\tabcolsep}{3pt}  
\begin{tabular}{|l|l|c|c|c|c|c|c|}
\hline
\textbf{Task} & \textbf{Model} & \textbf{Accuracy} & \textbf{Precision} & \textbf{F1-score} & \textbf{Explainability} & \textbf{Computational Cost} & \textbf{Preprocessing Dependence} \\ \hline
\multirow{4}{*}{Gender Classification} 
 & \textbf{TransECG (Ours)} & \textbf{0.899} & \textbf{0.900} & \textbf{0.899} & High (Attention-based) & High (ViT requires more computation) & High \\ \cline{2-8}
 & ECG Unveiled~\cite{wang2024ecg} & 0.755 & 0.766 & 0.760 & Medium (SHAP-based) & Low (Traditional ML) & Moderate \\ \cline{2-8}
 & CNN-based Model~\cite{ghazarian2021increased} & 0.857 & 0.860 & 0.855 & Low (Black-box) & Medium & Moderate \\ \cline{2-8}
 & LSTM-based Model~\cite{kim2020ecg} & 0.863 & 0.870 & 0.865 & Low (Black-box) & Medium-High (Sequential) & Moderate \\ \hline
\multirow{4}{*}{Age Group Classification}  
 & \textbf{TransECG (Ours)} & \textbf{0.899} & \textbf{0.901} & \textbf{0.899} & High (Attention-based) & High (ViT requires more computation) & High \\ \cline{2-8}
& ECG Unveiled & 0.671 & 0.623 & 0.633 & Medium (SHAP-based) & Low (Traditional ML) & Moderate \\ \cline{2-8}
 & CNN-based Model & 0.834 & 0.840 & 0.832 & Low (Black-box) & Medium & Moderate \\ \cline{2-8}
 & LSTM-based Model & 0.849 & 0.855 & 0.847 & Low (Black-box) & Medium-High (Sequential) & Moderate \\ \hline
\multirow{4}{*}{ID Re-identification} 
 & \textbf{TransECG (Ours)} & \textbf{0.886} & \textbf{0.889} & \textbf{0.882} & High (Attention-based) & High (ViT requires more computation) & High \\ \cline{2-8}
 & ECG Unveiled & 0.819 & 0.817 & 0.810 & Medium (SHAP-based) & Low (Traditional ML) & Moderate \\ \cline{2-8}
 & CNN-based Model & 0.842 & 0.845 & 0.840 & Low (Black-box) & Medium & Moderate \\ \cline{2-8}
 & LSTM-based Model & 0.855 & 0.858 & 0.852 & Low (Black-box) & Medium-High (Sequential) & Moderate \\ \hline
\end{tabular}
\label{tab:comparison}
\end{table*}

\subsection{Explainability for ECG Re-identification}

\subsubsection{Gender Classification} 

\begin{figure}[ht!]
    \centering
    \begin{subfigure}[b]{0.4\textwidth}
        \centering
        \includegraphics[width=\textwidth]
        {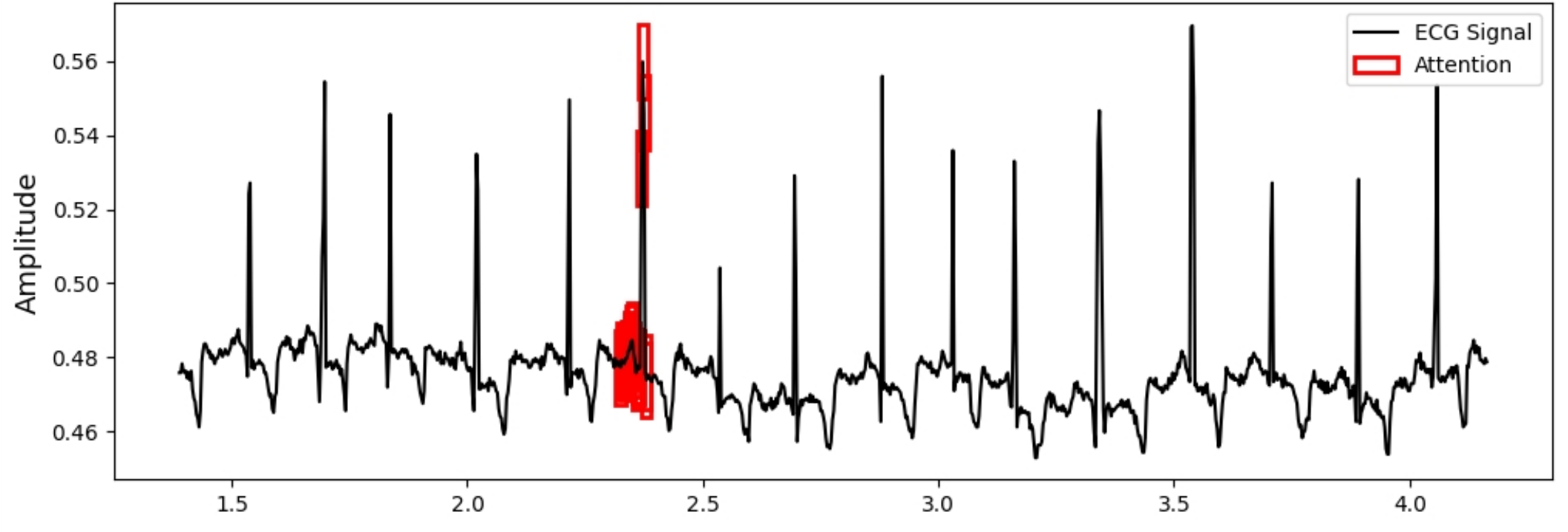}
        \caption{Head 1}
        \label{fig:head1}
    \end{subfigure}
    \hfill
    \begin{subfigure}[b]{0.4\textwidth}
        \centering
        \includegraphics[width=\textwidth]{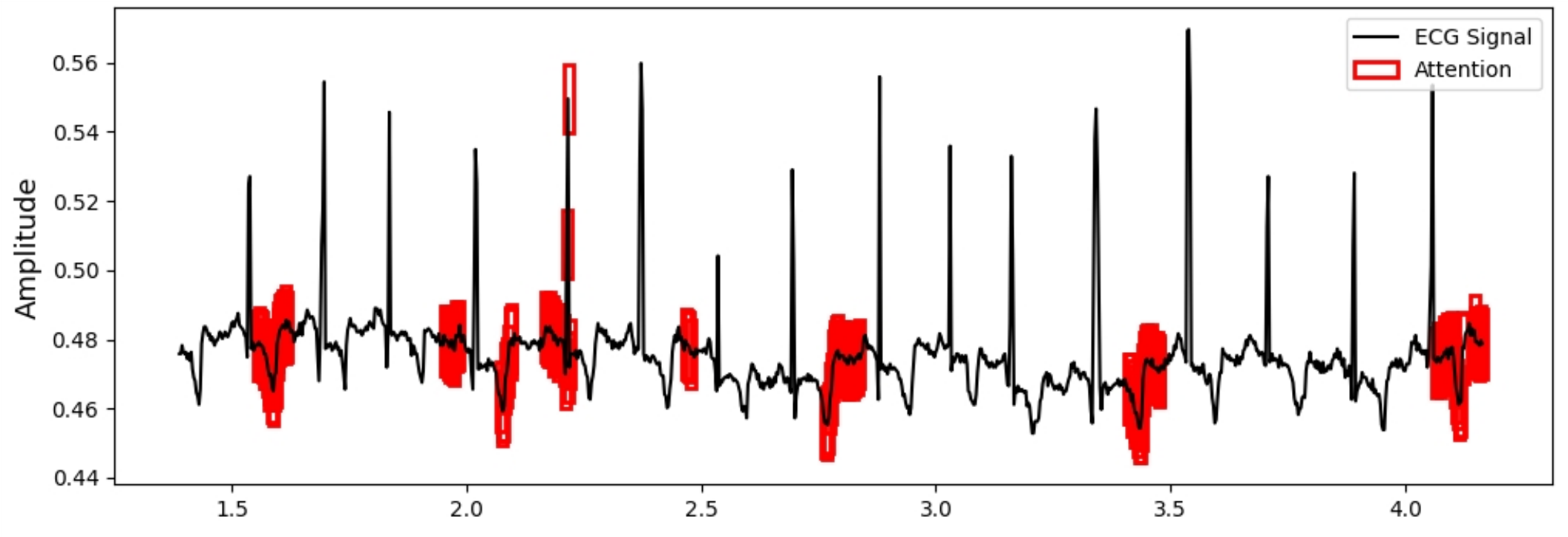}
        \caption{Head 2}
        \label{fig:head2}
    \end{subfigure}
    
    \begin{subfigure}[b]{0.4\textwidth}
        \centering
        \includegraphics[width=\textwidth]{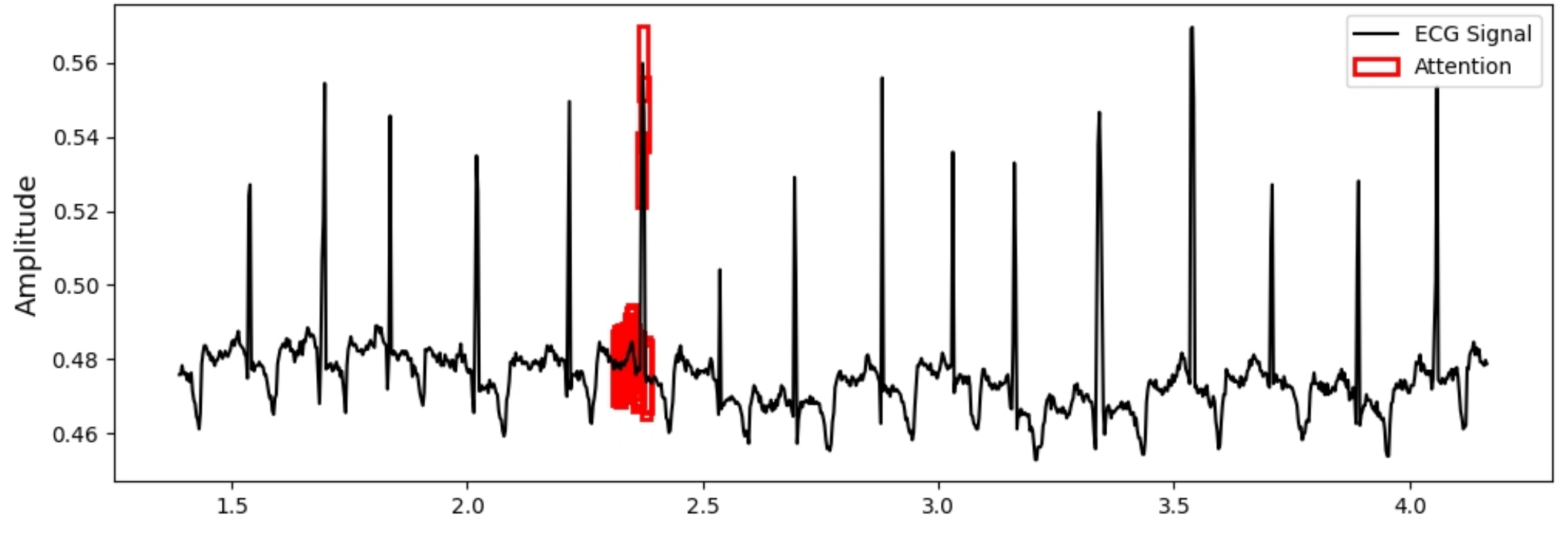}
        \caption{Head 3}
        \label{fig:head3}
    \end{subfigure}
    \hfill
    \begin{subfigure}[b]{0.4\textwidth}
        \centering
        \includegraphics[width=\textwidth]{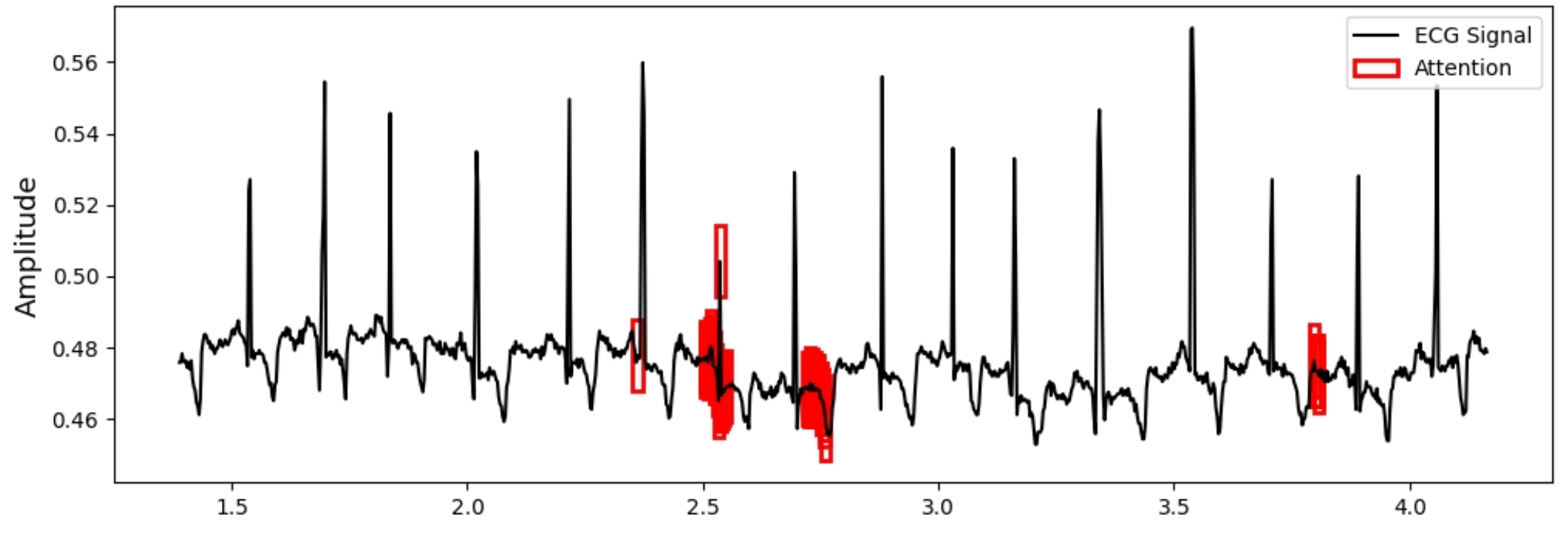}
        \caption{Head 4}
        \label{fig:head4}
    \end{subfigure}
    
    \begin{subfigure}[b]{0.4\textwidth}
        \centering
        \includegraphics[width=\textwidth]{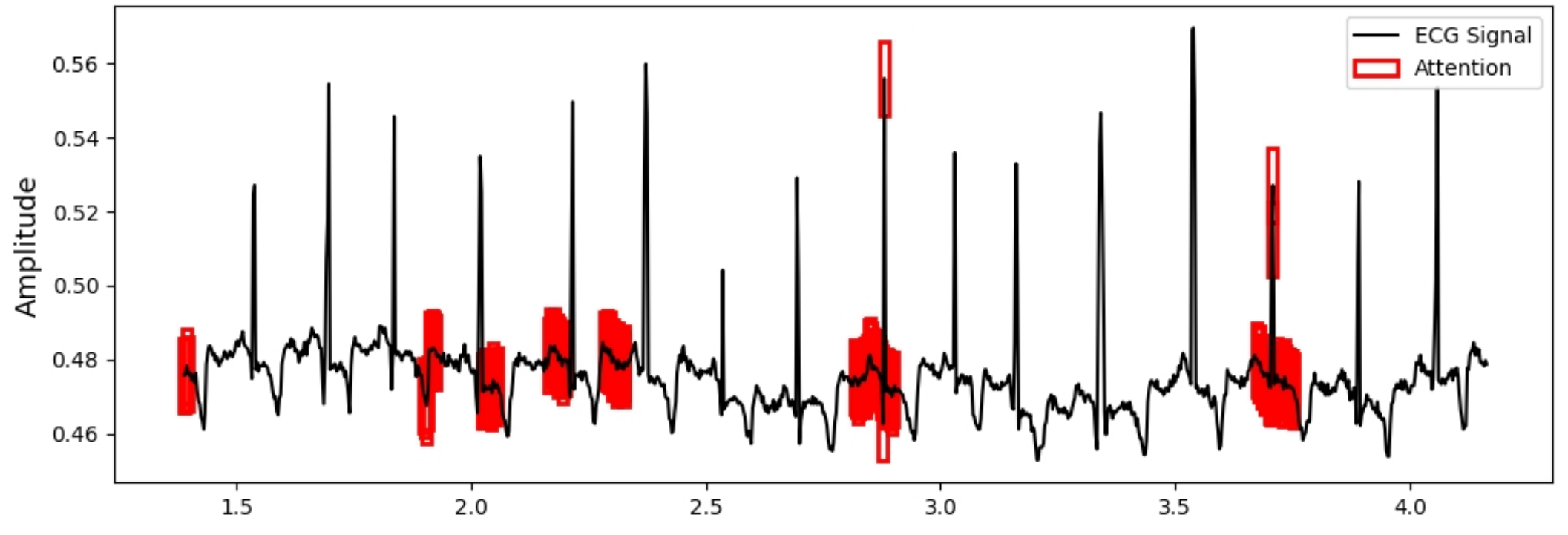}
        \caption{Head 5}
        \label{fig:head5}
    \end{subfigure}
    \hfill
    \begin{subfigure}[b]{0.4\textwidth}
        \centering
        \includegraphics[width=\textwidth]{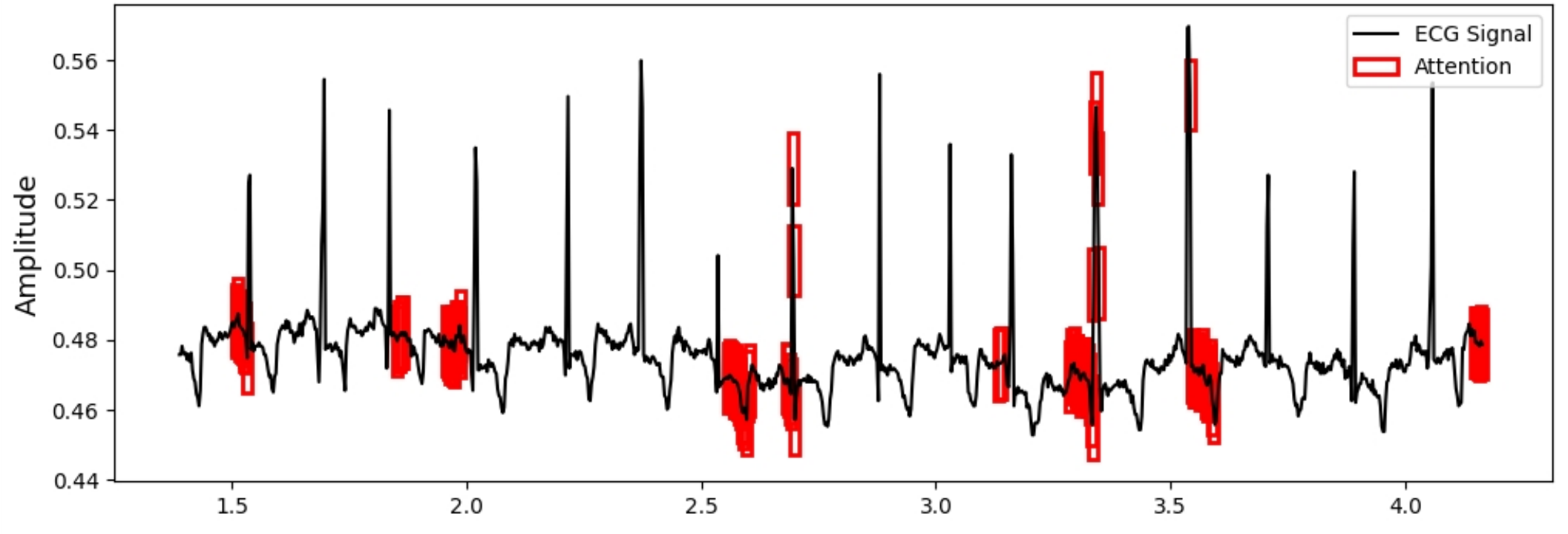}
        \caption{Head 6}
        \label{fig:head6}
    \end{subfigure}
    \caption{Gender Classification - Attention Maps}
    \label{fig:attention_maps}
\end{figure}

In this part, we analyzed attention scores captured from TransECG applied to one representative sample from ECG signals of test set for gender classification, focusing on the attention distribution across different components of the PQRST cycle. It is important to note that all these attention
heads correspond to the same ECG signal.  In the first attention head (Fig.~\ref{fig:head1}), the model assigns significant attention to the R peak, which corresponds to the P-R and R-T intervals. The R wave, representing ventricular depolarization, is critical for gender classification due to its higher amplitude in males, linked to larger heart muscle mass \cite{surawicz2003differences}. The second head (Fig.~\ref{fig:head2}) shifts attention to the S wave and the T wave, focusing on the S-T interval. This interval is vital for capturing ventricular repolarization, which exhibits notable gender differences, especially in the QT interval, as females generally have longer QT intervals, influenced by hormonal and physiological factors \cite{chandra2022sex}.

The third head (Fig.~\ref{fig:head3}) reinforces the importance of the R peak by focusing again on the QRS complex, confirming that ventricular depolarization is a key feature for distinguishing between genders \cite{aydin2016effects}. The fourth head (Fig.~\ref{fig:head4}) spreads its attention between the S wave and early T wave, particularly within the S-T interval, while also capturing part of the P-Q interval \cite{walsh2006electrocardiography, ayano2022interpretable}. This suggests that the model is considering atrial-to-ventricular conduction time in its classification process. The fifth head (Fig.~\ref{fig:head5}) expands attention across multiple intervals, including P-Q, R-S, and S-T, highlighting both conduction and repolarization phases. Similarly, the sixth head (Fig.~\ref{fig:head6}) distributes attention across the P-R, S-T, and T-Q intervals, emphasizing broader features of both depolarization and repolarization phases, reflecting subtle timing and amplitude variations between males and females \cite{wei2019physiology, james2007recent}.

Overall, TransECG demonstrates a consistent focus on medically significant features of the ECG signal, particularly the R wave, S-T interval, and P-Q interval. The attention on the R wave across multiple heads is validated by the fact that males generally exhibit larger R wave amplitudes due to greater ventricular mass. Meanwhile, the focus on the S-T interval aligns with known differences in repolarization, as females have longer QT intervals, making it an important feature for gender classification \cite{james2007recent}. The attention to the P-Q interval in several heads also suggests the model is capturing gender-related conduction differences between atrial and ventricular depolarization. Collectively, our TransECG model’s attention mechanism effectively captures amplitude and interval-based features that are crucial for gender classification, providing robust and explainable results.

We extract the projection matrix from the final block of the transformer model, enabling analysis of attention patterns across heads. The projection matrix is derived from the last layer of the encoder, where, after concatenating all the self-attention heads, a linear layer is applied to assign different weights to each head.  This helps identify which heads have been assigned higher weight at the final layer of encoder, consequently has higher contribution on the final classification task. For example, it shows us that heads 2 and 6 have higher attention score--1.0 and 0.933708, respectively. We also make a comparison among different sets of samples from our ECG dataset on the six heads--which reveals strong consistency in the model’s focus on the QRS complex, particularly the R-wave, across both samples, which is crucial for gender classification in ECG signals. Some heads consistently prioritize the R-wave, while others show a broader attention distribution across the PQRST cycle, capturing features from the P-wave, QRS complex, and T-wave in diverse analyses. However, there are slight variations, such as increased attention to the P-Q interval and T-wave in the different analysis (Figs.~\ref{fig:head1} to \ref{fig:head6}), which suggests that the model’s sensitivity to these smaller features may vary depending on the sample. Despite these minor differences, the model reliably highlights the most medically significant components of the ECG in different sets, indicating that its attention mechanisms are robust.

\subsubsection{Age Classification}
\begin{figure}[ht!]
    \centering
    \begin{subfigure}[b]{0.4\textwidth}
        \centering
        \includegraphics[width=\textwidth]{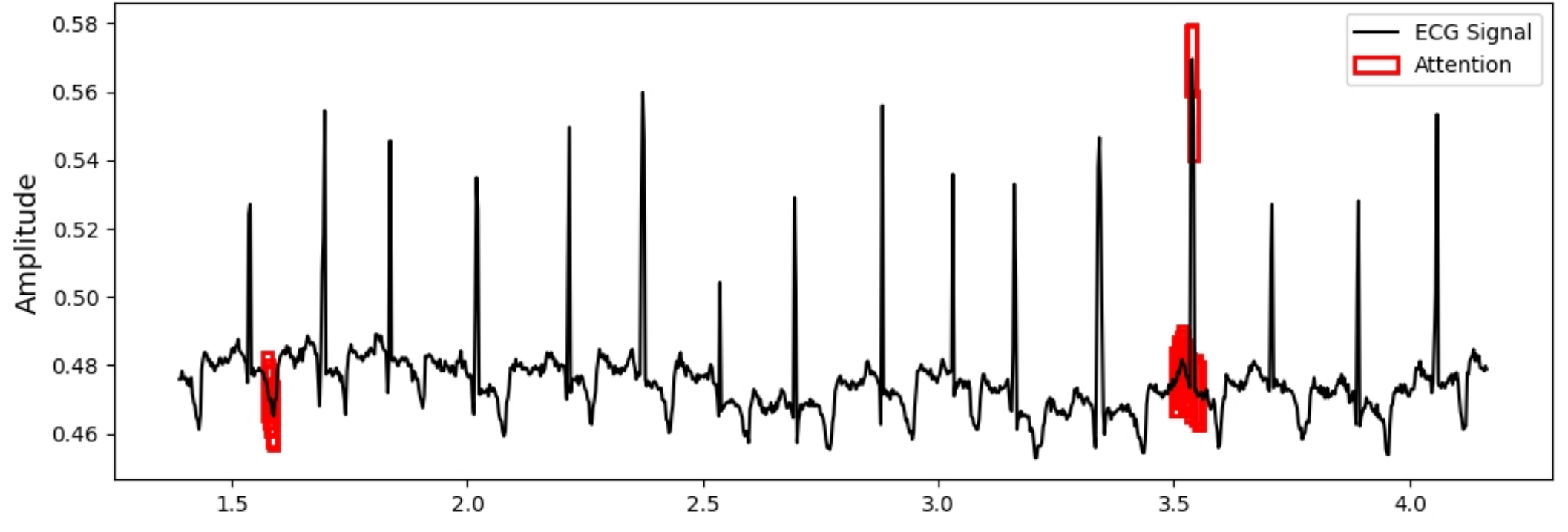}
        \caption{Head 1}
        \label{fig:headA1}
    \end{subfigure}
    \hfill
    \begin{subfigure}[b]{0.4\textwidth}
        \centering
        \includegraphics[width=\textwidth]{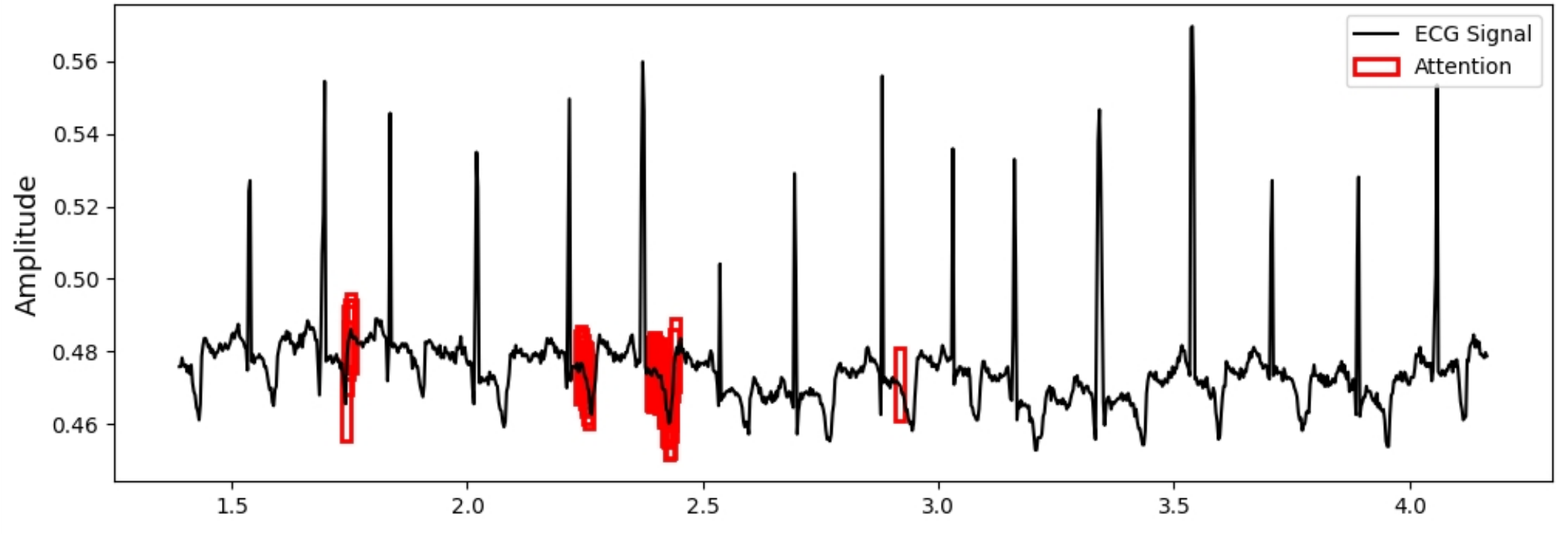}
        \caption{Head 2}
        \label{fig:headA2}
    \end{subfigure}
    
    \begin{subfigure}[b]{0.4\textwidth}
        \centering
        \includegraphics[width=\textwidth]{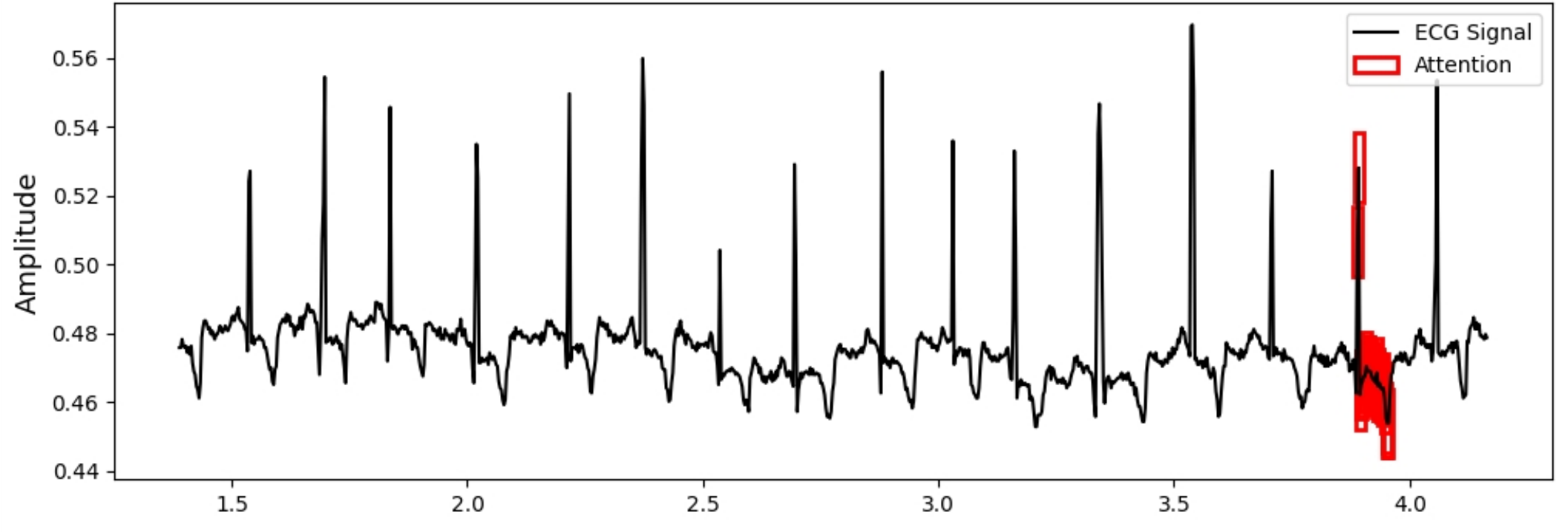}
        \caption{Head 3}
        \label{fig:headA3}
    \end{subfigure}
    \hfill
    \begin{subfigure}[b]{0.4\textwidth}
        \centering
        \includegraphics[width=\textwidth]{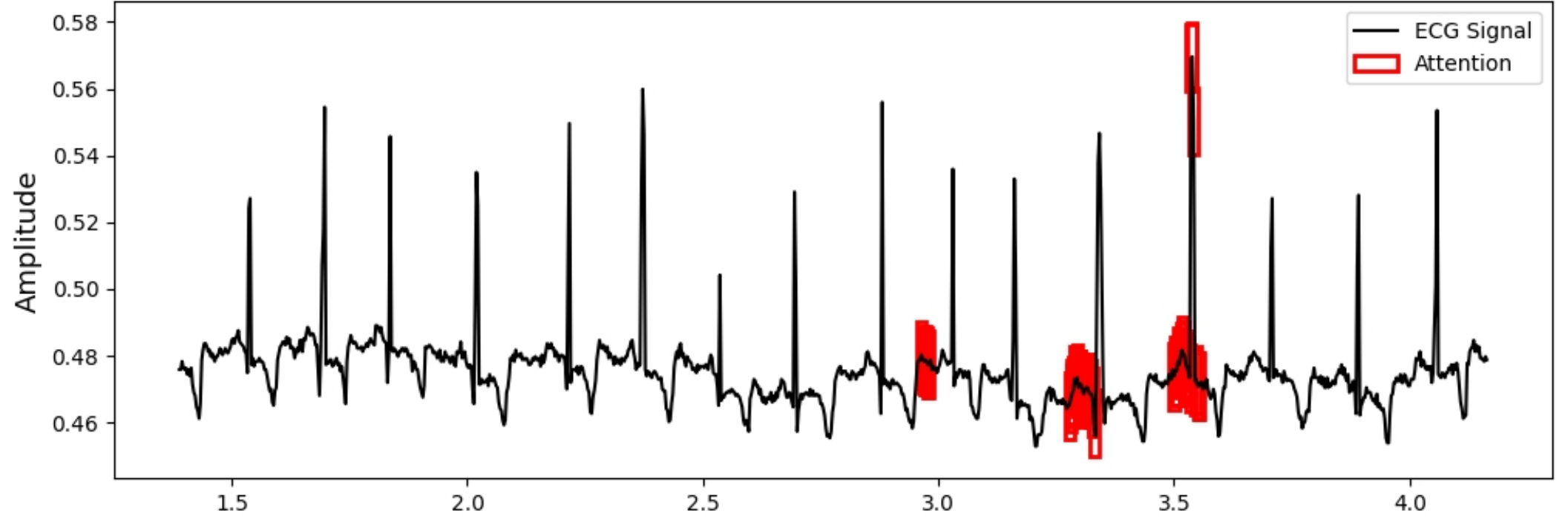}
        \caption{Head 4}
        \label{fig:headA4}
    \end{subfigure}
    
    \begin{subfigure}[b]{0.4\textwidth}
        \centering
        \includegraphics[width=\textwidth]{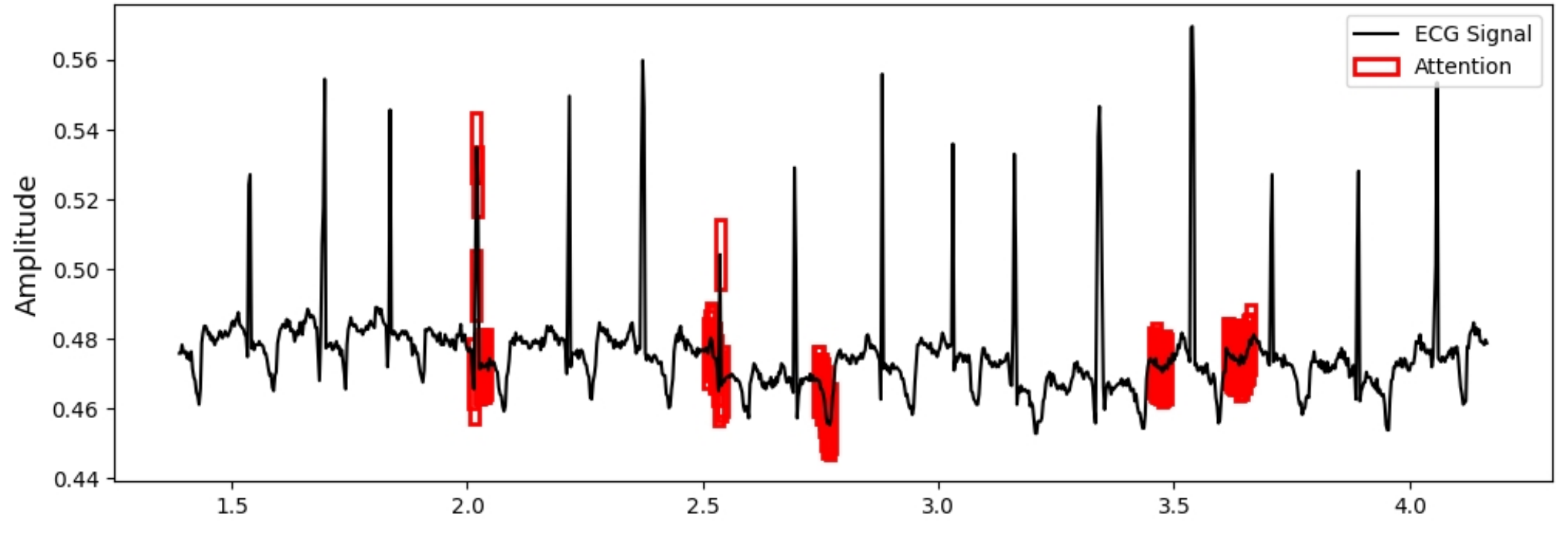}
        \caption{Head 5}
        \label{fig:headA5}
    \end{subfigure}
    \hfill
    \begin{subfigure}[b]{0.4\textwidth}
        \centering
        \includegraphics[width=\textwidth]{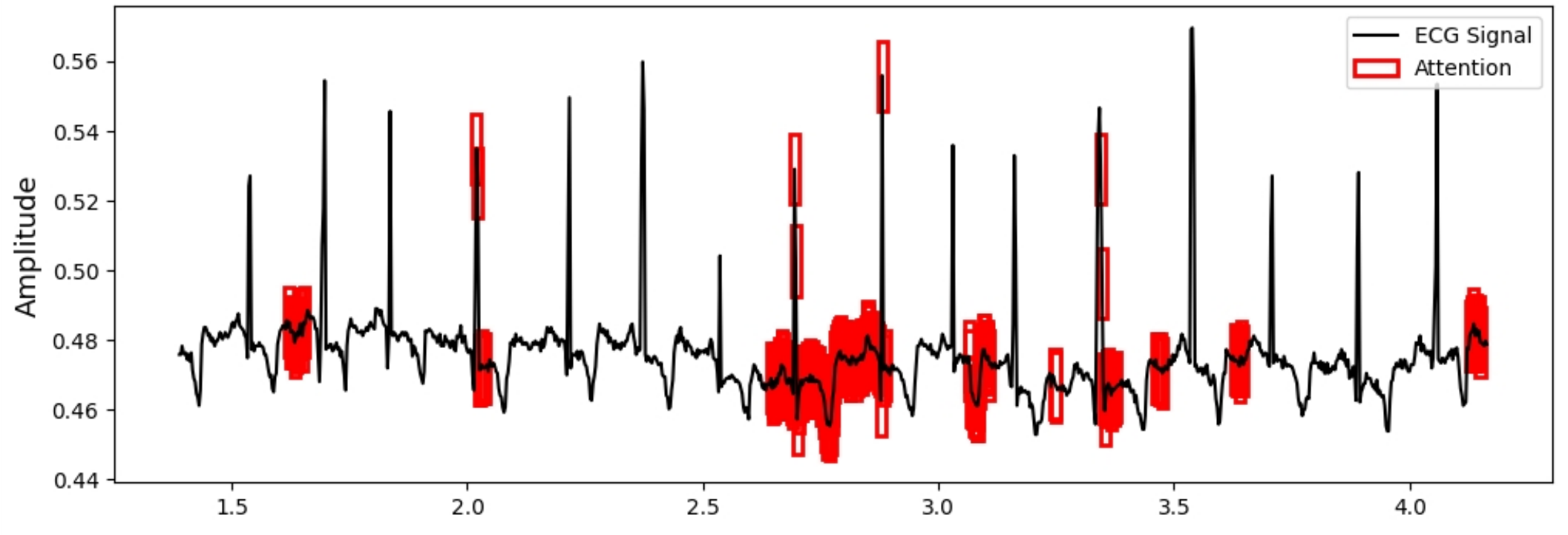}
        \caption{Head 6}
        \label{fig:headA6}
    \end{subfigure}
    \caption{Age Classification - Attention Maps}
    \label{fig:attention_maps_age}
\end{figure}

We analyzed attention maps generated by our TransECG model applied to one sample of ECG data for age classification, focusing on crucial components of the PQRST cycle. The attention visualization across six heads reveals distinct focus areas within the ECG signal for the age classification task. Head 1 (Fig.~\ref{fig:headA1}) emphasizes the P-R interval and R peak, highlighting the heart's depolarization phase, which is known to slow with age due to atrioventricular conduction delay \cite{swift2020age, kashou2017atrioventricular}. Head 2 (Fig.~\ref{fig:headA2}) centers on the S wave and T wave, focusing on ventricular repolarization, a critical aspect of cardiac function that can alter with age \cite{kenny2022ecg}. In Head 3 (Fig.~\ref{fig:headA3}), attention shifts to the P-R and Q-T intervals, indicating that both atrial depolarization and ventricular repolarization play key roles in the model’s decision-making \cite{sattar2023electrocardiogram}. Head 4 (Fig.~\ref{fig:headA4}) also focuses on the P-Q and S-T intervals, reflecting the model’s sensitivity to various phases of electrical activity, including repolarization and recovery phases \cite{kashou2017st}.

Heads 5 and 6 (Figs.~\ref{fig:headA5} and ~\ref{fig:headA6}) broaden the model’s focus to multiple intervals, particularly the Q-R, S-T, and P-T intervals, while continuing to emphasize the R and T wave amplitudes. These focus areas suggest that the model is not only tracking depolarization but also repolarization events, which are crucial for age-related classification due to the structural and electrical changes that occur in the heart with aging. The consistent attention to these features—specifically the P-R interval, Q-T interval, and R and T wave amplitudes—indicates that the model is capturing relevant clinical markers of aging, such as prolonged conduction times and changes in ventricular recovery \cite{surawicz2008chou, chadda2017effects}. Additionally, the projection matrix values for this sample show that heads 5 and 6 have higher contributions to the final classification task. These heads focus strongly on events such as the Q-R and S-T intervals, as well as the R and T wave amplitudes, which have higher attention scores, further underscoring their significance in age-related classification.

The model’s attention aligns well with known medical knowledge regarding cardiac aging. Prolongation of the P-R interval and alterations in the Q-T interval are established markers of aging, reflecting slower atrioventricular conduction and delayed ventricular recovery. The focus on R and T wave amplitudes is also validated by their association with age-related structural changes, such as increased heart mass and ventricular hypertrophy \cite{sidebotham2007cardiothoracic}. Overall, the model's attention effectively captures the key ECG features indicative of aging, making it a powerful tool for age classification in ECG analysis.

\subsubsection{Participant ID re-identification} 
\begin{figure}[ht!]
    \centering
    \begin{subfigure}[b]{0.4\textwidth}
        \centering
        \includegraphics[width=\textwidth]{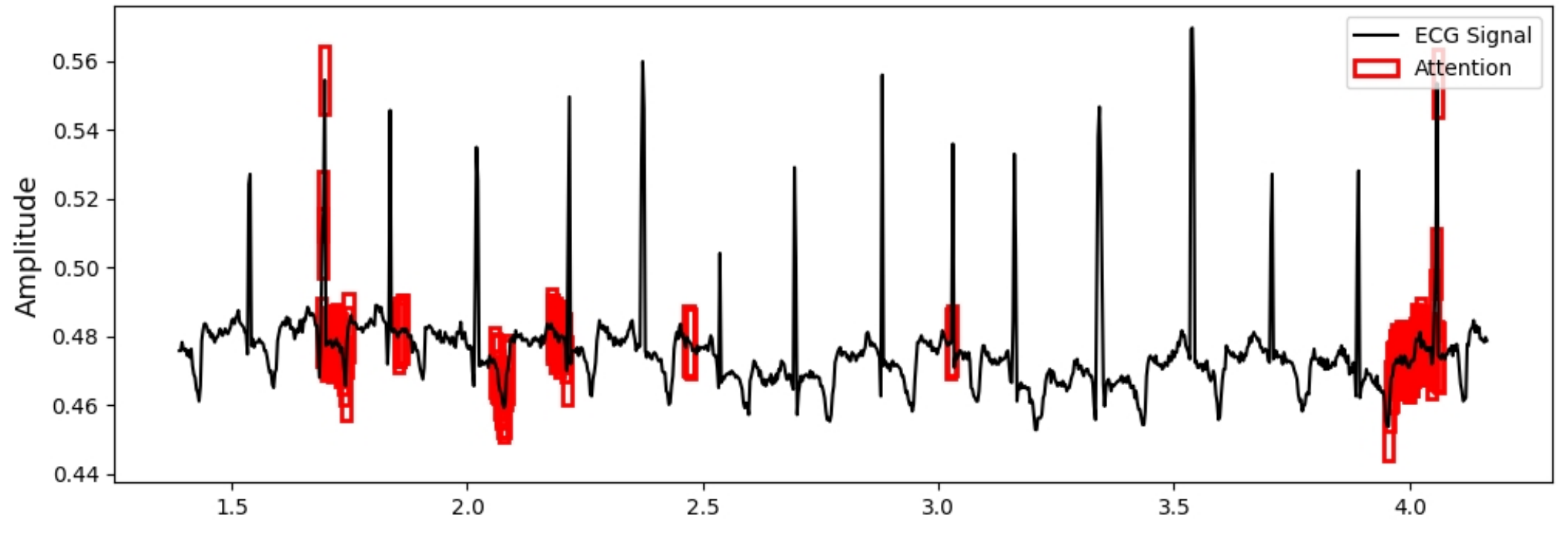}
        \caption{Head 1}
        \label{fig:headD1}
    \end{subfigure}
    \hfill
    \begin{subfigure}[b]{0.4\textwidth}
        \centering
        \includegraphics[width=\textwidth]{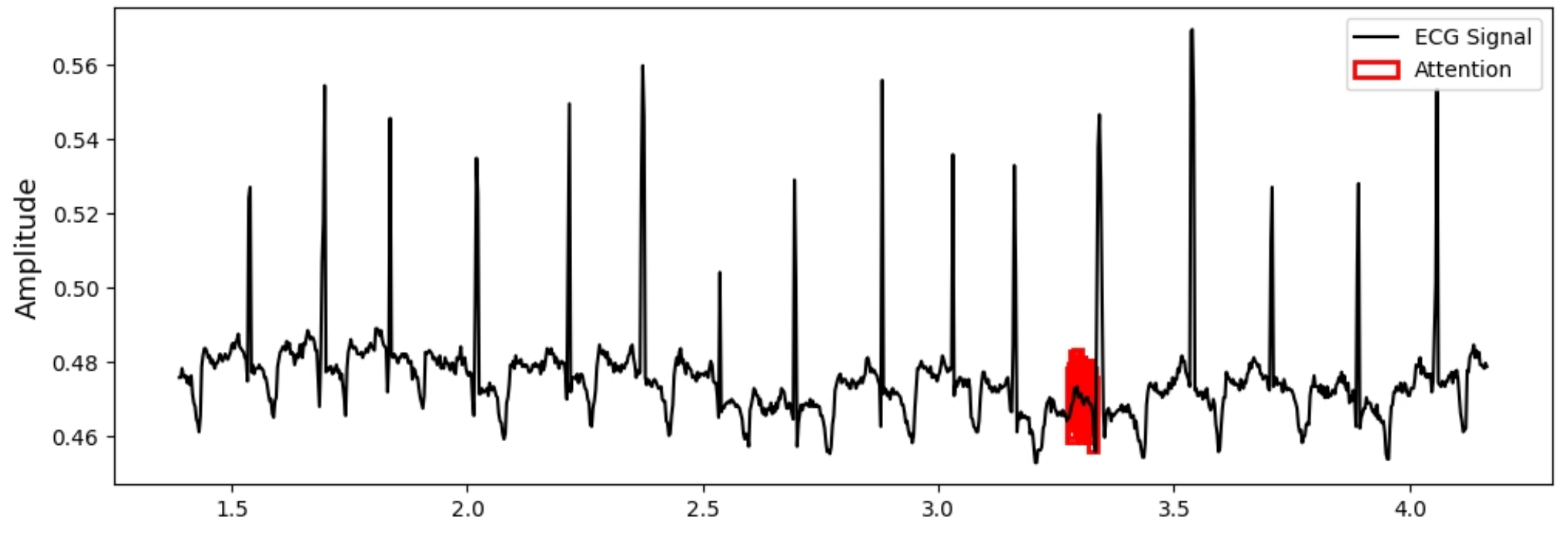}
        \caption{Head 2}
        \label{fig:headD2}
    \end{subfigure}
    
    \begin{subfigure}[b]{0.4\textwidth}
        \centering
        \includegraphics[width=\textwidth]{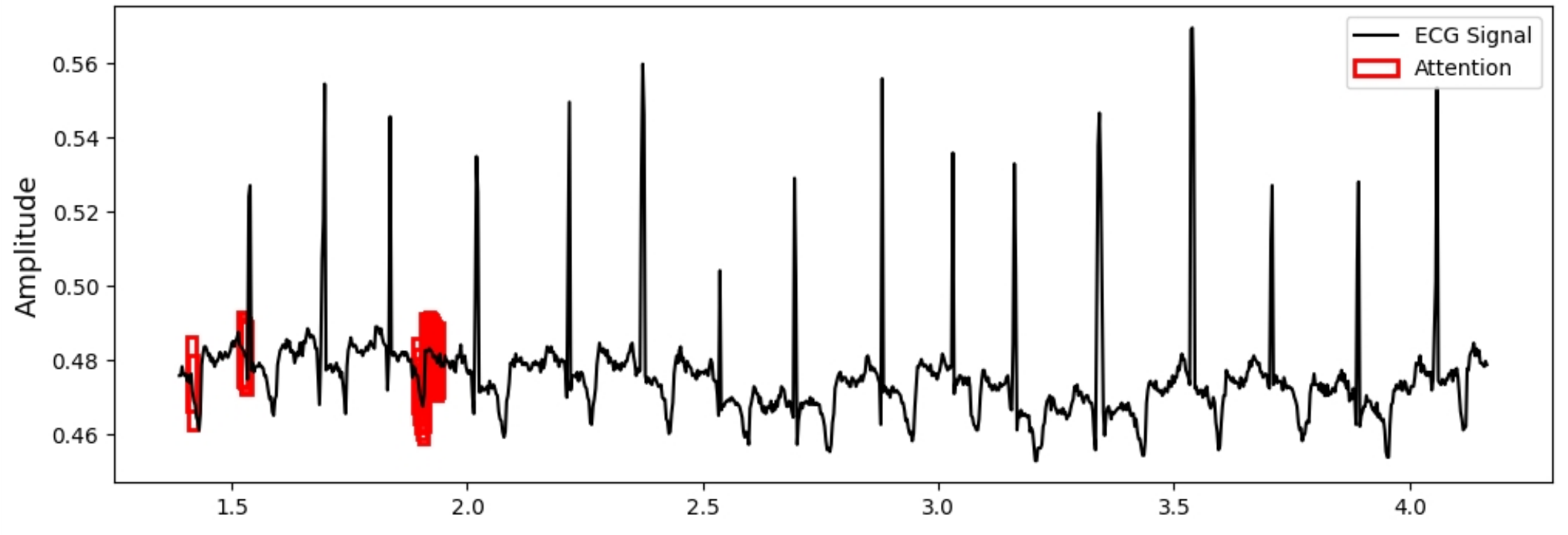}
        \caption{Head 3}
        \label{fig:headD3}
    \end{subfigure}
    \hfill
    \begin{subfigure}[b]{0.4\textwidth}
        \centering
        \includegraphics[width=\textwidth]{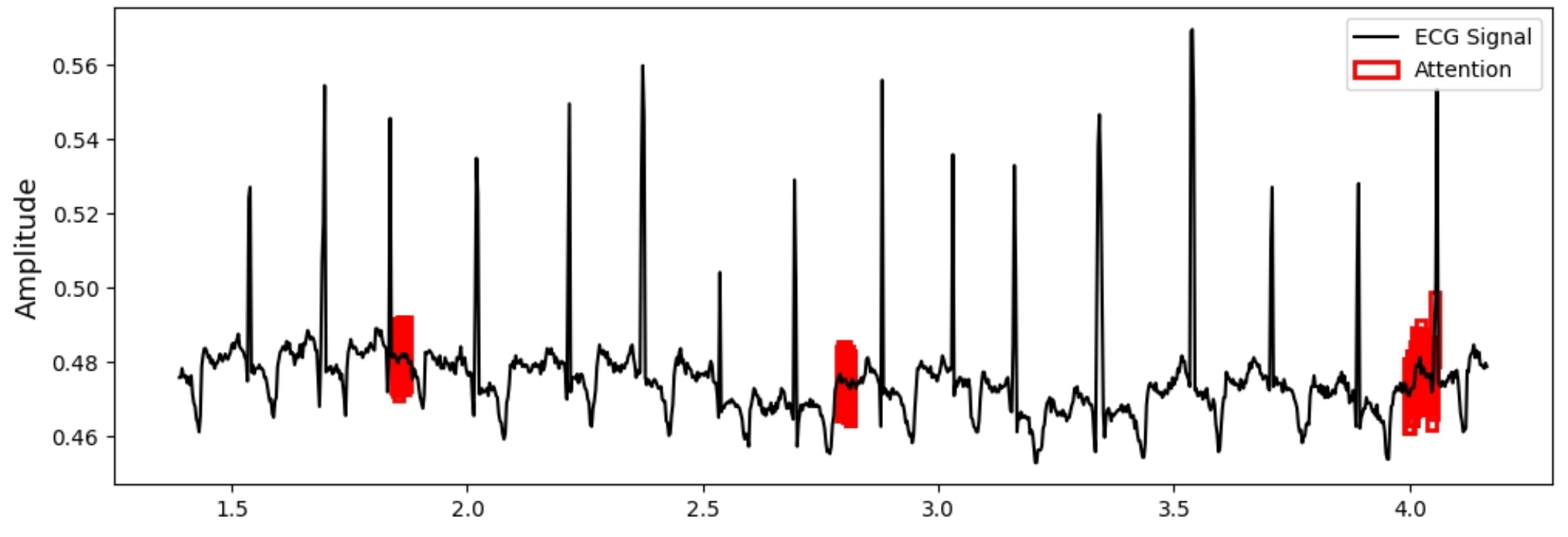}
        \caption{Head 4}
        \label{fig:headD4}
    \end{subfigure}
    
    \begin{subfigure}[b]{0.4\textwidth}
        \centering
        \includegraphics[width=\textwidth]{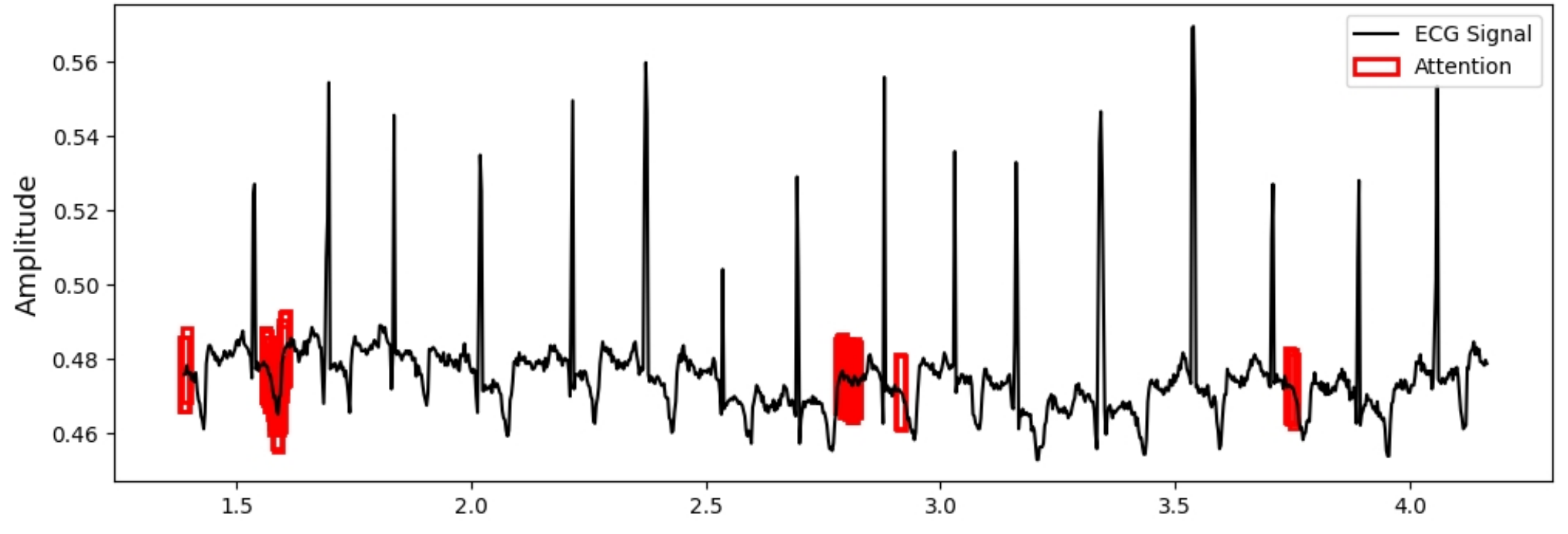}
        \caption{Head 5}
        \label{fig:headD5}
    \end{subfigure}
    \hfill
    \begin{subfigure}[b]{0.4\textwidth}
        \centering
        \includegraphics[width=\textwidth]{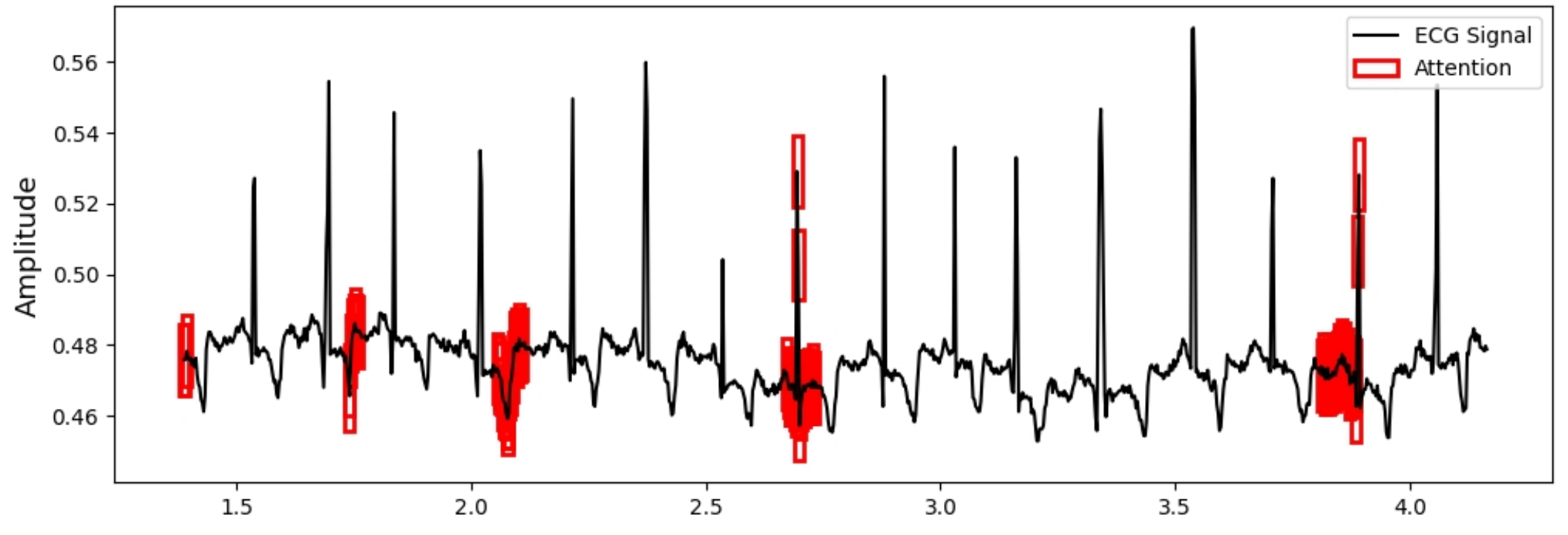}
        \caption{Head 6}
        \label{fig:headD6}
    \end{subfigure}
    \caption{ID re-Identification - Attention Maps}
    \label{fig:attention_maps_id}
\end{figure}

The analysis of the six heads in the last block of our TransECG model applied on one sample of ECG signals reveals distinct attention patterns focused on various ECG signal features. In Head 1 (Fig.~\ref{fig:headD1}), the attention is strongly centered around the R peak, highlighting the importance of the R-S and Q-R intervals in ventricular depolarization. The focus on the amplitude of the R wave, which typically indicates ventricular depolarization, suggests that individual variations in R-wave height due to heart size or electrical conduction differences are crucial for identification \cite{goldman2011goldman}. In Head 2 (Fig.~\ref{fig:headD2}), attention is focused on the S wave and T wave, emphasizing the S-T interval (ventricular repolarization) and T-Q interval (ventricular recovery) \cite{monitillo2016ventricular, hlaing2005ecg}. This suggests that the model captures repolarization dynamics that can vary significantly between individuals, providing vital cues for re-identification.

\begin{figure*}[!ht]
    \centering
    \includegraphics[width=0.93\linewidth]
    {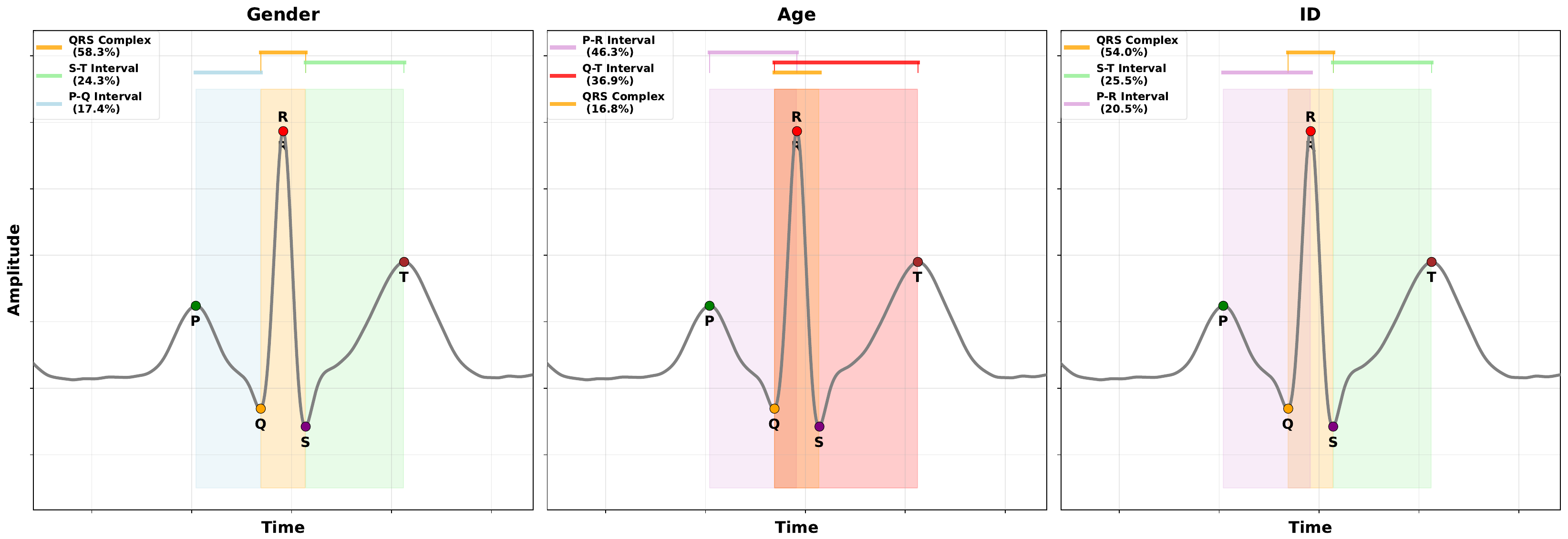}
    \caption{Visualization of ECG key features with attention scores for gender, age, and ID re-identification tasks. This figure complements Table~\ref{tab:ecg-features} by highlighting the regions of the ECG where the model concentrates, offering an interpretable representation of the ViT’s focus and insights for re-identification.}
    \label{fig:ecg-viz-attention}
\end{figure*}

Further analysis of Head 3 (Fig.~\ref{fig:headD3}) reveals attention on the P wave, particularly focusing on the P-R interval and P-T interval, reflecting atrial depolarization and the complete depolarization-repolarization cycle \cite{ashley2004cardiology}. This suggests that variations in atrial activity contribute to individual distinctions. Head 4 (Fig.~\ref{fig:headD4}) targets the Q and R waves, primarily focusing on the Q-R and Q-S intervals, capturing the duration and amplitude of ventricular depolarization \cite{feher2017quantitative}. This trend is further observed in Head 5 (Fig.~\ref{fig:headD5}), which emphasizes the P-Q and R-S intervals, indicating that both atrial depolarization and the spread of ventricular depolarization are key distinguishing features \cite{marr2011cardiology}. Head 6 (Fig.~\ref{fig:headD6}) spreads attention across the QRS complex and the T wave, highlighting the importance of both depolarization and repolarization dynamics \cite{ashley2004cardiology} for individual identification.

In conclusion, TransECG demonstrates a strong focus on the QRS complex, R wave, S wave, and T wave, indicating that variations in the ventricular depolarization and repolarization phases are critical for distinguishing individual ECG signals. The attention maps align with known medical knowledge, as these regions represent the electrical activity of the heart that varies between individuals due to anatomical and physiological differences. The focus on specific intervals (P-R, R-S, S-T) and amplitude features (R wave, T wave) confirms that these key components of the cardiac cycle are vital for the ID re-identification task. These findings are medically validated by existing literature on ECG signal analysis, where gender, heart size, and electrical conduction efficiency significantly influence the PQRST features.complete depolarization-repolarization cycle of the heart \cite{feher2017quantitative, ashley2004cardiology}.

\subsubsection{Attention-Driven Key ECG Patterns for Gender, Age, and ID re-identification}


This section highlights the top three key ECG features identified by the attention mechanism of TransECG for the tasks of gender identification, age classification, and ID re-identification.  The attention maps were generated from the last block of the Vision Transformer (ViT) model, focusing on the relationships between the class token and ECG patches. By normalizing attention scores and mapping them to the PQRST cycle, critical components such as the R-wave, P-R interval, and S-T segment were identified as highly influential for each task. These results were obtained by evaluating the model on the test dataset, extracting attention scores, and mapping them back to physiological features of the ECG signal. Further, the percentage contributions of these features were calculated by aggregating attention scores over the corresponding ECG intervals and normalizing them to reflect their relative importance--revealing the most influential features for each task as demonstrated in Table~\ref{tab:ecg-features}.

\begin{table}[!t]
\renewcommand{\arraystretch}{1.2}
\caption{ECG Key Features and Attention Percentage for Gender, Age, and ID Tasks}
\centering
\begin{tabular}{|l|c|c|c|}
\hline
\multirow{2}{*}{\textbf{Task}} & \multicolumn{3}{c|}{\textbf{ECG Key Features and Attention Percentage (\%)}} \\
\cline{2-4}
& \textbf{1st ECG Key} & \textbf{2nd ECG Key} & \textbf{3rd ECG Key} \\
& \textbf{Feature} & \textbf{Feature} & \textbf{Feature} \\
\hline
Gender & R-Wave (QRS & S-T Interval / & P-Q Interval / \\
& Complex) / 58.29\% & 24.28\% & 17.43\% \\
\hline
Age & P-R Interval / & Q-T Interval / & R-Wave (QRS \\
& 46.29\% & 36.87\% & Complex) / 16.84\% \\
\hline
ID & R-Wave (QRS & S-T Interval / & P-R Interval / \\
& Complex) / 53.97\% & 25.54\% & 20.49\% \\
\hline
\end{tabular}
\label{tab:ecg-features}
\end{table}

Table~\ref{tab:ecg-features} and Figure~\ref{fig:ecg-viz-attention} summarize the top three contributing ECG features for each task, along with the corresponding attention percentages. The attention percentages reflect the proportion of the model’s focus on each feature during classification, offering insights into the most critical ECG segments for each task. For gender identification, the model focuses heavily on the R-wave (QRS Complex), which receives 58.29\% of the attention, reflecting physiological differences in heart size and muscle mass between males and females. The S-T Interval receives 24.28\% and the P-Q Interval captures 17.43\%, further highlighting the model’s ability to capture gender-specific traits related to ventricular depolarization and repolarization.

In age classification, the model predominantly attends to the P-R Interval with 46.29\% of the focus, emphasizing its importance in distinguishing age-related changes in cardiac conduction. The Q-T Interval follows with 36.87\%, and the R-wave contributes 16.84\%, reflecting the model’s effectiveness in detecting subtle variations linked to age. For ID re-identification, the QRS Complex leads with 53.97\% of the attention, highlighting its significance in capturing individual-specific cardiac patterns. The model also emphasizes the S-T Interval with 25.54\% and the P-R Interval with 20.49\%, recognizing these features as key to personal identification through ECG signals. As illustrated in Figure~\ref{fig:ecg-viz-attention}, these findings confirm the clinical relevance of task-specific ECG features.

The ViT model’s attention scores highlight task-specific ECG features critical for classification. For gender, the model’s focus on the R-wave (58.29\%) aligns with known gender-based differences in ventricular depolarization. In age classification, the P-R interval (46.29\%) dominates, reflecting conduction delays associated with aging, while the Q-T interval (36.87\%) captures ventricular recovery changes. For ID re-identification, the R-wave (53.97\%) serves as a unique biometric marker, complemented by contributions from the S-T interval (25.54\%) and P-R interval (20.49\%). These findings are consistent with prior studies~\cite{surawicz2003differences}, confirming the clinical relevance of these features.

\subsubsection{ECG Features Extraction: Comparison of TransECG and ECG Unveiled
Models}

The ECG features extraction comparison between TransECG and ECG Unveiled highlights the significant advantages of using ViT and attention mechanisms. Unlike ECG Unveiled, which relies on manually crafted features like P-Q and R-S intervals, TransECG leverages attention mechanisms to capture complex and long-range dependencies, allowing for better identification of key ECG components such as the QRS complex, P-wave, and T-wave, which are crucial for tasks like gender and age classification.

TransECG can automatically identify important intervals like the S-T interval, which is often missed in manually crafted approaches. However, this interval is critical for reflecting the repolarization of ventricles, a key factor in re-identification tasks. By dynamically extracting these features, TransECG offers more detailed and flexible ECG analysis. Altogether, TransECG's attention-based method outperforms ECG Unveiled in tasks like gender classification, age group classification, and ID re-identification. The dynamic feature extraction in TransECG enhances both predictive performance and explainability, making it a superior choice for privacy-conscious healthcare applications.

\section{Discussion}

The results of this study demonstrate the dual strengths of TransECG: achieving high accuracy while providing task-specific explainability across gender, age, and participant ID classification tasks. The attention mechanisms highlighted critical ECG features such as the R-wave for gender identification, the P-R interval for age classification, and the QRS complex for ID re-identification. These task-specific insights not only align with established clinical knowledge but also validate the robustness of the model's interpretability, bridging the gap between predictive performance and transparency.

These findings have significant implications for real-world applications, particularly in enhancing data privacy. By identifying the most influential ECG features, targeted security algorithms can be developed to selectively add noise to sensitive regions, such as the R-wave or P-R interval, depending on the specific task. This task-specific obfuscation which we intend to follow as the future directions could help mitigate re-identification risks while preserving the analytical utility of the data. Such strategies are critical for enabling secure and ethical ECG data sharing, especially in healthcare environments.

Explainable and high-performance models like TransECG addresses the urgent need for privacy-conscious frameworks in biometric applications. These insights pave the way for advanced privacy-preserving techniques and support the development of trusted, transparent AI systems in healthcare.

\section{Conclusions}

In this paper, we introduced TransECG, a ViT-based method for ECG classification and re-identification risk analysis, leveraging attention mechanisms to enhance explainability and transparency in data sharing environments. Unlike existing methods that primarily focus on high identification accuracy, TransECG provides insights into the specific ECG features and intervals critical for identification tasks, addressing key privacy concerns. Our experiments demonstrate that TransECG achieves competitive accuracy across gender classification, age classification, and participant ID re-identification, with accuracy rates of 89.9\%, 89.9\%, and 88.6\%, respectively. Through attention analysis, we identified key ECG components like the R-wave and intervals such as P-R and S-T that significantly contribute to model decisions. For example, the R-wave accounted for 58.29\% of attention in gender classification, while the P-R interval dominated in age classification (46.29\%). These insights align with established medical knowledge, offering explainability in privacy-sensitive applications.

By combining predictive performance with interpretability,\\ TransECG offers a privacy-conscious solution for ECG data sharing, aligning with the goals of secure and trusted healthcare data spaces. Future work will incorporate additional XAI techniques to further enhance transparency and explore advanced privacy-preserving mechanisms tailored for data sharing frameworks.

\bibliographystyle{IEEEtran}
\bibliography{ref}
\end{document}